\documentclass[pdflatex,11pt]{article}

\usepackage{jheppub}
\usepackage{comment}
\usepackage{xcolor}
\usepackage{graphicx}
\usepackage{wrapfig}
\usepackage{multirow}
\usepackage{parskip}
\usepackage{tikz-cd}
\usepackage{chngcntr}
\counterwithout{figure}{section} 

\newcommand{\ali}[1]{\begin{align} #1 \end{align}}
\newcommand{\p}{\partial}

\newcommand{\ra}{\rightarrow}

\newcommand{\vev}[1]{\langle #1 \rangle} 
\newcommand{\mn}{{\mu\nu}}
\newcommand{\ab}{{\alpha\beta}}
\newcommand{\al}{\alpha}

\makeatletter
\gdef\@fpheader{}
\makeatother

\begin{document}
	\thispagestyle{empty}

	\title{Setting $T^2$ free for braneworld holography} 
	
	\author{Nele Callebaut}\author{and Matteo Selle}
	\affiliation{Institute for Theoretical Physics, University of Cologne, Z\"{u}lpicher Stra\ss e 77, 50937 K\"{o}ln, Germany} 
	
	\abstract{ 
		We identify what has been referred to as `cut-off CFT' in holographic braneworld with $T^2$ or $T\bar T$ theory (depending on the dimension of the bulk), so that the holographic dual of AdS-gravity with Neumann boundary conditions is a $T^2$-deformed CFT that is set free. After making
		statements that apply for general dimensions higher than three, we focus on the case of a three-dimensional bulk. 
		We find from bulk arguments that the effective theory on the brane is governed by a $T\bar T$-like flow equation, such that under certain assumptions the effective gravity theory on the brane is given by a $T\bar T$-like deformed timelike Liouville theory, which limits to the description of the holographic Weyl anomaly for branes that approach the asymptotic boundary. 
	}
	
	\maketitle

	\title{Setting $T^2$ free for braneworld holography} 
	
	\section{Introduction and overview of results}

Braneworld holography has a long history. The famous work on braneworld by Randall and Sundrum \cite{Randall:1999ee,Randall:1999vf}  
was interpreted holographically in seminal works by Gubser \cite{Gubser:1999vj} and Verlinde \cite{Verlinde:1999fy}, and further investigated by many authors, such as in \cite{Giddings:2000mu,Karch:2000ct,Verlinde:1999xm,Porrati:2001gx,Perez-Victoria:2001lex,Arkani-Hamed:2000ijo}. 
The main idea was that bulk gravity with Neumann boundary conditions (NBC) or `braneworld' has a holographic dual interpretation as the dual CFT of the Dirichlet boundary condition (DBC) problem coupled to an effective gravity theory on the brane.  

The use of braneworld theories in holography was revived in recent years, often going under the name of double holography  \cite{Chen:2020uac,Karch:2000ct,Geng:2022slq,Geng:2022tfc,Neuenfeld:2024gta} or AdS/bCFT \cite{Takayanagi:2011zk,Fujita:2011fp,Suzuki:2022xwv}, 
particularly in the context of the black hole information paradox. 
To model a version of the paradox, toy models involving braneworld set-ups were constructed and used  successfully  to derive the `island formula' for the entropy of black hole radiation,  
which essentially resolves the long-standing paradox \cite{Almheiri:2019hni,Almheiri:2019yqk,Chen:2020uac,Penington:2019npb,Penington:2019kki,Almheiri:2019qdq,Verheijden:2021yrb,Almheiri:2020cfm,Hartman:2020swn}. In \cite{Almheiri:2019hni}, the bulk theory is $3D$ AdS-gravity with part of its boundary the regular asymptotic boundary with DBC, and another part of its boundary given by a so-called end-of-the-world (EOW) brane, with NBC.  
This divides the dual boundary model into a region that has just the CFT and a region with CFT matter coupled to gravity, according to respectively regular AdS/CFT holography and braneworld holography.   
Typically, however, the effective gravity theory on the brane is introduced by hand to be a particular model, e.g.~Jackiw-Teitelboim (JT) gravity, in order to investigate the black hole info paradox in that particular $2d$  gravity set-up.

Now the question is what the effective gravity theory on the brane should be from a bulk gravity integration calculation. Results in this direction have been reported e.g.~in \cite{Geng:2022slq,Geng:2022tfc,Deng:2022yll,Aguilar-Gutierrez:2023tic}, on retrieving JT gravity, and \cite{Neuenfeld:2024gta}, discussing  
a Liouville gravity theory. Their holographic considerations were in terms of an effective gravity theory coupled to the dual CFT or ``cut-off CFT'', accounting for the brane being away from the asymptotic boundary.     
In this work, we set out to address the question of deriving the braneworld gravity theory for branes at any finite radial location in the bulk, and 
without restricting a priori to small fluctuations.  
That is, we want to investigate the \emph{general} NBC problem in AdS-gravity and more precisely its holographic dual.

\begin{figure}[t]
	\centering
	\includegraphics[scale=0.35]{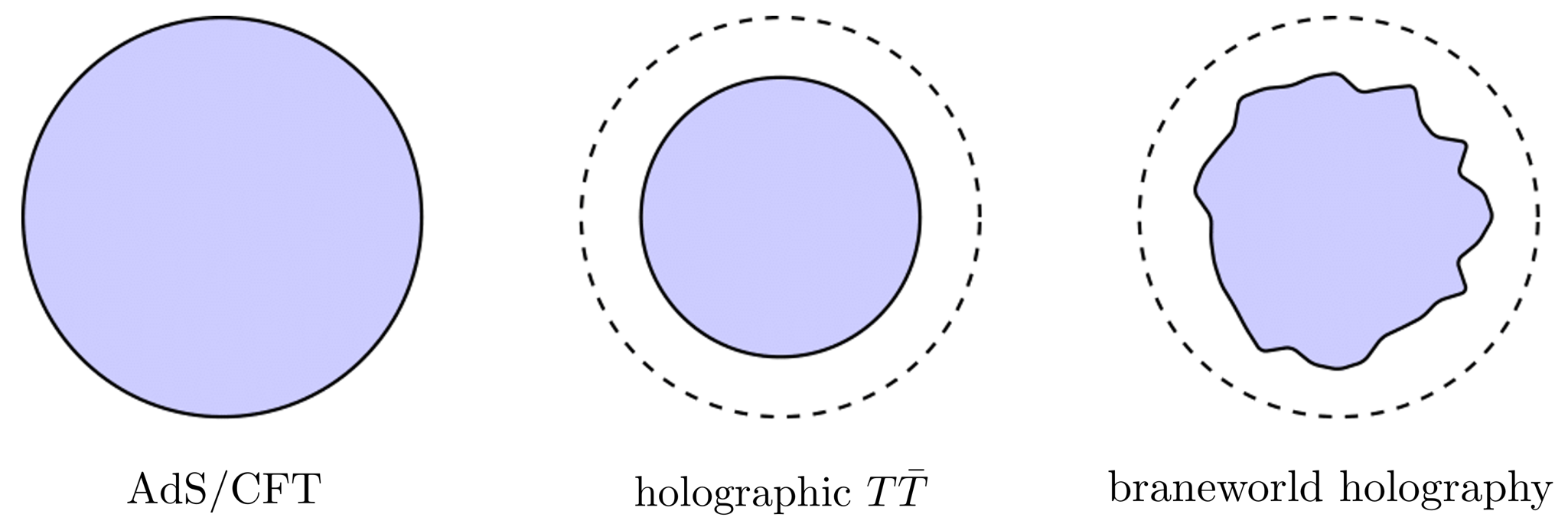}
	\caption{A schematic representation of different relevant types of holography: AdS/CFT, $T\bar T$ holography or `cut-off holography' and finally braneworld holography. They correspond respectively to asymptotic DBC, DBC at finite cut-off and NBC in AdS-gravity. The wiggly contour in the representation of braneworld holography indicates that NBC do not fix the boundary geometry.} \label{figholo}
\end{figure}

In other recent developments, there have been different avenues exploring holography beyond standard AdS/CFT, as ultimately one wishes to understand non-AdS (dS or flat) quantum gravity. One is the use of different boundary conditions. This 
includes besides DBC and NBC the so-far unmentioned conformal boundary conditions (CBC) \cite{Witten:2022xxp,Allameh:2025gsa,Anninos:2023epi,Coleman:2020jte} and mixed boundary conditions (MBC) \cite{Guica:2019nzm, Bzowski:2018pcy}.  
Another is `finite' holography: it was discovered in \cite{McGough:2016lol}  and further investigated in e.g.~\cite{Kraus:2018xrn,Belin:2020oib,WallAraujo-Regado:2022gvw,Caputa:2020lpa,Mazenc:2019cfg,Apolo:2023vnm} that in the case of pure $3D$ AdS-gravity,  imposing DBC at a finite distance into the bulk  
corresponds to deforming the dual CFT with a particular operator called the $T\bar T$ operator. It gives rise to a $T\bar T$-deformed CFT or in short $T\bar T$ theory, or to a $T^2$ theory in higher-dimensional set-ups \cite{Hartman:2018tkw,Taylor:2018xcy}.   
This is commonly referred to as $T\bar T$ holography or `cut-off holography', as the bulk is cut off at a finite radial location. Given this name, it is not surprising that what is called `cut-off CFT' in braneworld holography will indeed be identified in section \ref{sect1} with the $T\bar T$ theory.  
The different types of holography that play a role in our discussion are illustrated in Fig.~\ref{figholo}.  

Other works that investigate the interplay between $T\bar T$ and braneworld holography are \cite{Ondo:2022zgf,Deng:2023pjs,Kawamoto:2023wzj,Basu:2024xjq,Parvizi:2025wsg,Hirano:2025cjg}.  

\paragraph{Overview} 
This brings us to an overview of the paper and its main results. We start in section \ref{sect1} by reviewing the standard holographic braneworld argument in subsection \ref{subsect1-1}, and updating it in terms of $T\bar T$ language in section \ref{subsect1-2}. 
The general statement on the holographic interpretation of the braneworld theory $Z_{bw}$ is given in \eqref{bwholo}, in terms of a \emph{$T^2$-deformed} CFT coupled to an effective braneworld gravity theory, which is given explicitly in \eqref{Sbwgravd}. Next, we consider $3D$ bulk gravity in particular, for the main part of the paper. It needs to be considered separately, as explained in the introduction of section \ref{sectWeyl}, and is the main case of interest for us, 
having in mind the island set-ups as motivation and the holographic description of $T\bar T$ being best understood in this number of dimensions. 

The discussion of $3D/2d$ braneworld holography is split into two sections. In section \ref{sectWeyl} we restrict to asymptotic branes, for which the dual interpretation will be in terms of the CFT  
that is set free (adopting the language of \cite{Compere:2008us}). 
Much of this section can be considered a review and we compare explicitly to the literature, but it sets our notation and serves as the basis for our later extensions into the bulk. 
The boundary perspective is addressed first, in subsection \ref{subs2d}: this contains a review of the integrated Weyl anomaly in $2d$ CFT, $Z_{CFT}(g_{(0)}) = Z_{CFT}(\hat g_{(0)}) \, \exp\{i S_L[\phi]\}$,   
which introduces the Liouville field and action $S_L[\phi]$ that will play an important role in the whole paper.  
We work with a `triangle representation' of a conformal transformation \eqref{bdytriangle1} or `boundary triangle' 
and pause at the role of the Liouville stress tensor. 
The Liouville action $S_L[\phi]$ is then derived from a bulk calculation in subsection \ref{subs3d},
essentially following \cite{Carlip:2005tz,Takayanagi:2018pml} but expanding on the method. It is a rederivation of the holographic Weyl anomaly, where we work with \emph{finite} $\phi$ specifically. 
The set-up of the calculation heavily relies on the bulk interpretation of the `boundary triangle' in terms of a triangle representation of Brown-Henneaux diffeomorphisms \eqref{bulktriangle} or `bulk triangle'. Namely, it clarifies the equivalence between calculating the difference between  on-shell actions along the diagonal arrow in the bulk triangle, \eqref{FGmhatFGrefined}, and the difference along the vertical arrow, \eqref{hatFGmhatFG}. This is illustrated in Fig.~\ref{trianglecirclesfigure}. The result is indeed the (timelike) Liouville action (with vanishing cosmological constant) in \eqref{SLholoWeyl}, for $W_{CFT}(g_{(0)}) - W_{CFT}(\hat g_{(0)})$. We compare to the extensive, original work on the holographic Weyl anomaly by Skenderis and collaborators, and point out a side result of the understood equivalence (between \eqref{FGmhatFGrefined} and \eqref{hatFGmhatFG}): the asymptotic Weyl mode $\phi$ can also be thought of as describing the physics of a Fefferman-Graham horizon. This is illustrated in Fig.~\ref{figbulkhorizon}. In subsection \ref{subsholo} on asymptotic braneworld, we set the obtained integrated Weyl anomaly free. Here we discuss why we effectively only integrate over $\phi$ at large central charge, and we address how the $S_L$ Liouville theory differs from other, effective Liouville theories that appear in AdS$_3$-gravity.  

Next, in section \ref{sectTT}, we move on to the main part of the paper, which is to consider branes at a general radial location in the bulk, and setting $T\bar T$ free in order to obtain a prescription for the braneworld holography theory by $Z_{bw} = \int D g Z_{T\bar T}(g)$. We restrict our considerations here to braneworld with flat reference metric $\hat g$ or unit tension $T=1$, which 
for comparison to traditional braneworld corresponds to a Randall-Sundrum type choice of flat slicing in the bulk. In subsection \ref{subsectTTWeyl}, we simply extend the strategy of the bulk calculation of section \ref{sectWeyl}   
further into the bulk, calculating the difference of on-shell actions $W_{T\bar T}(g) - W_{T\bar T}(\hat g)$ as presented in Fig.~\ref{trianglecirclesfigure}, in practice along the vertical arrow. 
The result for the action $S_{\tilde L} \equiv W_{T\bar T}(g) - W_{T\bar T}(\hat g)$ (defined as the difference) 
is a higher-derivative extension of the timelike Liouville action $S_L$, given in \eqref{StildeLtildephiresult}, \eqref{StildeLPT} and \eqref{StildeLphiresult} in different incarnations. These are the main results of this section. The expression in \eqref{StildeLtildephiresult} is in terms of a newly introduced field $\tilde \phi$, which measures the position-dependent or `wiggly' location of the boundary at $\rho = \bar \rho + \tilde \phi(x)$ 
in the top right corner of the bulk triangle in  Fig.~\ref{trianglecirclesfigure}. It is the equivalence between the diagonal and vertical arrow in that figure that explains the connections between setting $T\bar T$ free and $3D$ braneworld calculations (e.g.~in \cite{Geng:2022slq,Geng:2022tfc,Neuenfeld:2024gta}) that let the brane fluctuate $\bar \rho \ra \bar \rho + \tilde \phi(x)$. Since the whole picture represents the Brown-Henneaux diffeomorphisms labeled by $\phi$, it schematically clarifies the relation between $\phi$ and $\tilde \phi$, and thus between 
the different approaches in terms of different fields.  

In section \ref{subsectTTSL} we take a different approach to setting $T\bar T$ free and consider the gravitational Hamilton-Jacobi equation in the bulk, which tells us precisely how the on-shell bulk actions depend on the induced metric $g$ of the boundary or their Weyl mode $\sigma$. This leads to a trace flow equation for the stress tensor $t_{\tilde L}^\mn$ of $S_{\tilde L}$, given in \eqref{traceflowliou}. 
The corresponding flow equation for the action $S_{\tilde L}\equiv S_{\tilde L}^{(t)}$ is the `$T\bar T$-like flow'   
\ali{
	\frac{d}{dt} S_{\tilde L}^{(t)} &= \frac{1}{4\pi} \int d^2 x \sqrt{-\hat g} \,  e^{-\sigma} \mathcal O^{\tilde L}_{T\bar T,\hat g}, \qquad S_{\tilde L}^{(0)} = S_L \label{actionflow0} 
}
copied from \eqref{actionflow}, with $\mathcal O^{\tilde L}_{T\bar T,\hat g} \equiv t_{\tilde L}^\mn t_{\tilde L}^\ab \hat g_{\alpha\mu} \hat g_{\beta\nu} - (t_{\tilde L}^\mn \hat g_\mn)^2$. It is characterized by a deformation that includes a Weyl factor, $t \ra t e^{-\sigma}$, compared to a $T\bar T$ flow. We discuss
the solution \eqref{StildeLsigmat1} (to first order in the deformation parameter $t$) and the solution \eqref{StildeLsigma} to the simpler problem where $\hat \Box \sigma = 0$, with seed theory the free boson. Finally, in section \ref{subsectTTfree}, the action $S_{\tilde L}$ takes on the interpretation of effective gravity action on the brane when we set $T\bar T$ free, 
\ali{
	Z_{bw} &= Z_{T\bar T}(\hat g) \int \mathcal D \sigma \,  e^{i S_{\tilde L}[\sigma]} , 
} 
for large central charge. 
In the limit of small $\sigma$ but for general cut-off radius,  
our braneworld holography reduces to the $T\bar T$-deformed CFT coupled to (timelike) Liouville theory with zero cosmological constant (equivalent to our choice of $T=1$). 
We end the paper with a comparison to the traditional braneworld strategy in section \ref{sectbw}, followed by a discussion and outlook in section \ref{sectdiscussion}. 

While preparing this manuscript, the paper \cite{Allameh:2025gsa} appeared. It has some overlap with the results we present, in that a flow equation is discussed that is of the same $T\bar T$-like ($t \ra t e^{-\sigma}$) type \eqref{actionflow0} as discussed in this work. 
The context in that paper is different, namely the CBC problem. However, a similar discussion as for the NBC problem is expected to some extent, because in $3D/2d$ the Weyl mode is also the only physical mode of the boundary metric (locally). We leave a better understanding of the connections to  
this recent paper for future work. Other work matched the $T\bar T$-deformed Liouville theory to a bulk holographic interpretation \cite{FarajiAstaneh:2024fig}.

	\section{$D$-dimensional braneworld holography} \label{sect1}
	
	Let us start by repeating the standard holographic interpretation of braneworld theories \cite{Gubser:1999vj,Giddings:2000mu,Verlinde:1999fy},  
	following in particular \cite{Gubser:1999vj,Giddings:2000mu}. 
	Before combining them into such an interpretation, we outline the standard set-ups of AdS/CFT holography and of braneworld. 
	This will set our notation for the  
	central ideas we will be building on. 
	
	Some words on notation. Throughout, we will use capital $G$ with capital indices for the $D = (d+1)$-dimensional bulk metric $G_{MN}$, and lower case $g$ with Greek indices for the $d$-dimensional boundary  metric $g_\mn$. The explicit indices on the metric fields will often be left out for clarity.  
	Riemann curvature tensors will be indexed by the metric they are calculated for in case of ambiguity, but we reserve $R$ for the Riemann curvature of the $d$-dimensional metric $g$. The scale of AdS-gravity is set by the AdS radius $l$. We will work in Lorentzian signature with mostly plus convention.

	\subsection{Traditional braneworld holography in terms of `cut-off CFT'} \label{subsect1-1}
	
	\paragraph{Holography}	
	In AdS/CFT holography, the central object is the gravitational path integral 
	\ali{
		Z_{grav}(g_{(0)}) &= \int_{G_{\tilde \p} \,=\, g_{(0)}} DG \,e^{i S_{grav}[G]}, \qquad  S_{grav} = S_{EH} + S_{GH} + S_{ct}.  \label{Zgrav}
		   	}
	It depends on $g_{(0)}$ as Dirichlet boundary condition input to the path integral. In this context, Dirichlet boundary conditions or AdS$_{d+1}$-gravity boundary conditions mean that one fixes the conformal boundary, denoted as $G_{\tilde \p}$, to $g_{(0)}$. In practice, the action is integrated up to an infinitesimal
	regulator distance $\epsilon$ away from the asymptotic boundary, where the induced metric $G_{\p} \equiv g$ is imposed to blow up as $G_{\p} = \frac{l^2}{\epsilon^2} g_{(0)}$. 
	The boundary condition  
	in this way fixes the first term in a Fefferman-Graham (FG) expansion of any asymptotically AdS metric  
	\ali{
		ds^2 &= G_{MN} dX^M dX^N  
		= \frac{l^2}{r^2} dr^2 + \gamma_\mn(x,r) dx^\mu dx^\nu  \\ 
		\gamma(x,r) &= \frac{l^2}{r^2}  \left( g_{(0)}(x) + r^2 g_{(2)}(x) + \cdots + r^d g_{(d)}(x) + h_{(d)}(x) r^d \log r^2 + \mathcal O(r^{d+1}) \right). \label{gind} 
	}  
On-shell, the higher order functions $g_{(2),(4),...}$ are 
determined in terms of $g_{(0)}$ and  $g_{(d)}$\footnote{The function $g_{(0)}$ determines recursively all higher order functions $g_{(2)}$, ..., up to $g_{(d-2)}$, as well as the trace and covariant divergence of $g_{(d)}$. 
Once $g_{(0)}$ and the unconstrained part of $g_{(d)}$ are specified, the entire FG expansion is completely fixed by the bulk Einstein's equations \cite{deHaro:2000vlm}.}. We reserve the notation $g$ for the induced metric, which is equal to $\gamma$ only for hypersurfaces at \emph{constant} $r$.

	The gravitational action $S_{grav}$ consists of the Einstein-Hilbert and Gibbons-Hawking-York contributions 
	\ali{
		S_{EH} + S_{GH} = \frac{1}{2\kappa_D} \int_{\mathcal M} 
		d^{D} X \sqrt{-G} \left( R_G  - 2 \Lambda_D \right) 
		+\frac{1}{\kappa_D} \int_{\p \mathcal M} 
		d^d x \sqrt{-g} \, K 
	}
	with $\Lambda_{D} = -\frac{d(d-1)}{2l^2}$, and counterterms $S_{ct}$. These are constructed to cancel the divergent terms in the on-shell action as $\epsilon$ is taken to zero, up to one $\log \epsilon$ contribution,   
	\ali{ 
		S_{ct} = -S^{div}_{EH + GH}[G_*(g_{(0)})] + \frac{1}{2\kappa_D} \int_{\partial \mathcal{M}}d^dx \sqrt{-g_{(0)}} \log \epsilon \,  a_{(d)}      \label{Sct}
	} 
	with $a_{(d)}$ a local covariant expression of $g_{(0)}$ which vanishes for odd $d$ \cite{Henningson:1998gx,deHaro:2000vlm}.   
	The resulting counterterms depend on $g_{(0)}$ and the cut-off $\epsilon$, but when written in terms of the induced metric $g$ no longer depend explicitly on $\epsilon$ \footnote{The holographic counterterms can be fully determined in terms of $g_{(0)}$ precisely because 
	the power-law divergent terms depend only on the 
	modes $g_{(0)},g_{(2)}, ..., g_{(d-2)}$.}.  
	Hence, we will write $S_{ct}[g]$, and $S_{grav} = S_{EH}[G] + S_{GH}[g] + S_{ct}[g]$. 
	The $\log \epsilon$ term represents the only contribution to the divergent action that would have introduced an explicit cut-off dependence in $S_{ct}$. 
	By instead retaining the $\log \epsilon$ divergence in the on-shell  $S_{grav}$, the explicit cut-off dependence takes on the physical interpretation of UV cut-off dependence in the CFT. Indeed, 
	consistent with this notion of holographic RG \cite{deBoer:1999tgo}, the coefficient $a_{(d)}$ is identified 
	as the Weyl anomaly of the dual CFT in \cite{Henningson:1998gx,deHaro:2000vlm}. 
	What we call $S_{ct}$ here is the counterterm action of Balasubramanian and Kraus in \cite{Balasubramanian:1999re}, compared to the counterterm action $S_{ct}^{Sk} \equiv -S^{div}_{EH + GH}[G_*(g_{(0)})]$ of Skenderis et.~al.~in \cite{deHaro:2000vlm,Skenderis:2000in,Skenderis:2002wp}. For completeness \cite{deHaro:2000vlm,Emparan:1999pm},   
\ali{
	S_{ct} &= -\frac{1}{\kappa_{d+1}} \int_{\p \mathcal M} d^d x \sqrt{-g} \Bigg[(d-1) + \frac{1}{2(d-2)}R  \nonumber \\ 
	&\qquad  + \frac{1}{2(d-4)(d-2)^2}\left(R_\mn R^\mn - \frac{d}{4(d-1)}R^2\right) + \cdots \Bigg]. 
}

	The variation of the action takes the form  
	\ali{
		\delta S_{grav} = \int_{\mathcal M} d^D X (EOM)\delta G + \frac{1}{2} \int_{\p \mathcal M} d^d x\sqrt{-g} \, T^\mn_{BY} \delta g_\mn  \label{BYdef}
	} 
	with the boundary contribution providing the definition of the Brown-York stress tensor $T^\mn_{BY}$\footnote{We are using here the convention $T^{\mu\nu}_{BY}=\frac{2}{\sqrt{-g}}\frac{\delta S_{grav}}{\delta g_{\mu\nu}}$. Later, when we specify to $D=3$, we will adopt the convention $T^{\mu\nu}_{BY}=\frac{4\pi}{\sqrt{-g}}\frac{\delta S_{grav}}{\delta g_{\mu\nu}}$.}. It is given by $T^\mn_{BY} = -\frac{1}{\kappa_D} \left( K^\mn - K g^\mn \right) + 2 \delta S_{ct}/(\sqrt{-g}\delta g_\mn)$ which we will write as 
	\ali{
		T^\mn_{BY} = T^\mn_{BY \text{w/o ct}} + T^\mn_{BY,\text{ ct contrib}}. \label{TBY}
	} 
	The Dirichlet variational problem, fixing the induced metric $\delta g_\mn=0$ at $r=\epsilon$ as described above, is thus well-defined and imposes the bulk EOM for $G_{MN}$.   
	
	\paragraph{Braneworld}  	
	Next we turn to braneworld, where the set-up is only slightly different in that the gravitational action contains a tension term $S_T$ for the brane rather than a boundary counterterm $S_{ct}$. We will accordingly use a different notation for the `total' action,  
	\ali{
		S_{tot}
		= S_{EH}[G] + S_{GH}[g] + S_T[g], \qquad S_T = -\frac{1}{\kappa_D} \, T \int \sqrt{-g}  
	} 
	to distinguish it explicitly from $S_{grav}$ in \eqref{Zgrav}. 
	Rather than fixing the bulk metric at a regulated  
	boundary, spacetime is bounded by a brane with an induced metric $g_\mn$ that is allowed to be dynamical, hence describing induced $d$-dimensional gravity on the brane. That is, one considers the Neumann variational problem for the variation 
	\ali{
		\delta S_{tot} = \int_{\mathcal M} d^D X (EOM)\delta G - \frac{1}{2 \kappa_D} \int_{\p \mathcal M} d^d x \sqrt{-g} \, \left(K^\mn - K g^\mn + T g^\mn \right) \delta g_\mn \label{deltaStot}
	}
	which requires setting the boundary expression multiplying $\delta g_\mn$ to zero. 
	We write the braneworld theory as 
	\ali{
		Z_{bw} = \int_{NBC} DG \, e^{i S_{tot}[G]}, \qquad \text{Neumann bc: } \quad K_\mn - K g_\mn + T g_\mn = 0 .  \label{Neumann}
	}
	To make contact to 
	the gravitational theory $Z_{grav}$ in
	AdS/CFT within the semi-classical limit, one can think of the braneworld construction as a two-step process. 
	First, constructing $Z_{tot}$ by integrating over bulk metrics that satisfy a Dirichlet boundary condition at the brane $G_\p = g$, and then 
	integrating over all possible values of $g$:  
	\ali{
		Z_{tot}(g) &= \int_{G_\p \, = \, g} DG \, e^{i S_{tot}[G]}, \qquad S_{tot} = S_{EH} + S_{GH} + S_T   \label{Ztot} \\  
		Z_{bw} &= \int Dg \, Z_{tot}(g).  \label{Zbw}
	}  
	The NBC are enforced in the saddle-point limit by the path integral over $g$ \cite{Giddings:2000mu,Compere:2008us}. 
	The 
	step of integrating over $g$ is referred to as `setting the boundary free' in \cite{Compere:2008us}, and we will be using the same terminology.  
	The `boundary' in this case is the brane located at a constant value $r = \bar r$ of the FG coordinate. This constant value is not required to be small, or said otherwise, the brane is not required to be close to the asymptotic boundary. When it is, we will refer to it as a \emph{near-boundary} brane. 
	In  
	braneworld, one typically considers a saddle point evaluation  
	\ali{
		Z_{bw} \approx \int Dh \, Z_{tot}(\hat g + h) \label{Zbwsaddle}
	}
	for $\hat g$ the saddle satisfying the Neumann boundary conditions \eqref{Neumann} and $h$ a small fluctuation, with the effective gravitational action on the brane quadratic in $h$. 
In the original literature  (e.g.~\cite{Randall:1999ee,Randall:1999vf,Karch:2000ct}), braneworld constructions often involve the gluing of two bulk spacetimes along one or more branes, with the imposition of Israel's junction conditions or orbifold boundary conditions. We will be following the more modern bottom-up braneworld constructions, in which a single bulk geometry is cut off by so-called end-of-the-world (EOW) branes with Neumann boundary conditions (e.g.~\cite{Chen:2020uac,Geng:2022slq,Geng:2022tfc,Neuenfeld:2024gta}). The two constructions should be related by orbifold symmetry. We will compare to braneworld literature in some more detail in section \ref{sectbw}. 

	The path integrals with Dirichlet boundary conditions in braneworld and holographic theories, respectively \eqref{Ztot} 
	and 
	\ali{
		Z_{grav}(g) &= \int_{G_\p = g} DG \, e^{i S_{grav}[G]}, \qquad S_{grav} = S_{EH} + S_{GH} + S_{ct}  \label{Zgravofg}
	}
	are simply related as 
	\ali{
		Z_{tot}(g) = Z_{grav}(g) e^{i (S_T[g] - S_{ct}[g])}  \label{ZtottoZgrav}
	}
	where we made use of the fact that the tension and counter-terms are boundary terms. 
	Setting $g$ free subsequently to obtain the braneworld theory $Z_{bw}$, will give a $d$-dimensional gravity interpretation to the terms in the exponent, which we therefore call `braneworld gravity' 
	\ali{
		S_{bwgrav} \equiv S_T - S_{ct}. \label{Sbwgravdef}
	} 
	Eqs.~\eqref{ZtottoZgrav} and \eqref{Sbwgravdef} form the basis for the holographic interpretation of braneworld $Z_{bw}$, given that we can build on the holographic interpretation of $Z_{grav}$. 
	
	The AdS/CFT correspondence can be succinctly stated as the equivalence between the gravitational path integral \eqref{Zgrav} with fixed conformal boundary $g_{(0)}$ and the $d$-dimensional CFT path integral $Z_{CFT}$ that depends on $g_{(0)}$ as a source i.e.~background metric\footnote{The remaining gravitational free data, consisting of the 
			unconstrained part of $g_{(d)}$, corresponds in Lorentzian signature 
			to a choice of state for the dual CFT (i.e. expectation value of the CFT's stress tensor).}, 
	\ali{
		Z_{grav}(g_{(0)}) &= Z_{CFT}(g_{(0)}) \qquad \text{(AdS/CFT)}. \label{AdSCFTregular}
	}   
	The object appearing in the braneworld discussion via \eqref{ZtottoZgrav} is instead $Z_{grav}(g)$, defined separately in \eqref{Zgravofg}. The difference between $Z_{grav}(g_{(0)})$ in \eqref{Zgrav} and $Z_{grav}(g)$ in \eqref{Zgravofg} is the following. In the former, the Dirichlet condition fixes the conformal boundary $G_{\tilde \p} = g_{(0)}$ or $G_\p = \frac{l^2}{\epsilon^2} g_{(0)} + \mathcal O(1)$ for  
	$\epsilon/l$ infinitesimally small. In the latter, the Dirichlet condition fixes the induced metric $G_\p = g$ at a general location 
	in the bulk. It is this  
	difference that in the literature is captured by introducing the terminology ``cut-off CFT'', implicitly referring to the location $\epsilon/l$ being treated as a small parameter in a perturbative FG expansion. In a modified version of AdS/CFT, it could be written as 
	\ali{
		Z_{grav}\left(g =  \frac{l^2}{\epsilon^2} g_{(0)} + \mathcal O(1)
		\right) = Z_{\text{``cut-off CFT''}}(g)    
		\qquad \text{(AdS/``cut-off CFT'')}   \label{AdScutoff}
	}
for a CFT with a 
	cut-off imposed at an energy scale that corresponds to the theory living at a boundary a distance $\epsilon$ into the bulk. The statement \eqref{AdScutoff} constitutes the definition of the concept ``cut-off CFT'', with $Z_{grav}$ on the left hand side including the same counterterms $S_{ct}$ as used asymptotically in regular AdS/CFT \eqref{AdSCFTregular}, as is natural in 
	this precise context of small $\epsilon/l$ FG-expansion. 	
Here, the relation between induced metric and conformal boundary needs to be systematically corrected to the FG expansion expression $g = g(x,\epsilon)$ in \eqref{gind}. 
	This means we can apply  
	this holographic duality to the case of 
a near-boundary	brane, to arrive (using \eqref{Zbw}) at the holographic interpretation of braneworld \cite{Gubser:1999vj,Giddings:2000mu} as   
	\ali{
		Z_{bw} &= \int Dg \, Z_{\text{``cut-off CFT''}}(g) e^{i S_{bwgrav}[g]}. 
	}
	It describes the coupling of the dual ``cut-off CFT'' to the 
	effective braneworld gravity \eqref{Sbwgravdef} as the dual interpretation to bulk gravity bounded by a near-boundary (EOW) brane  
	\ali{
		Z_{bw} &= \int Dg \, e^{i\left( W_{\text{``cut-off CFT''}}[g]+ S_{bwgrav}[g]\right)} 
		\qquad \text{(near-boundary braneworld holo)}. \label{bwholooriginal}
	}
	
	Now, for a \emph{general} brane, at any distance $r = \bar r$ in the bulk, we need a holographic interpretation of $Z_{grav}(g)$ in \eqref{Zgravofg}. Gubser in \cite{Gubser:1999vj} does in fact consider finite $\bar r$ and imposes the Dirichlet boundary condition in a perturbative expansion away from the boundary, for large $l/\bar r$, using the FG expansion $G_\p = g(x,\bar r)$ as a derivative expansion,  with $g(x,\bar r)$ to be read off from \eqref{gind}.  
	He still refers to this as a ``cut-off CFT'' 
	and to the holographic interpretation of braneworld as \eqref{bwholooriginal}. 
	But in modern parlance, it is in fact none other than the $T^2$-deformed CFT. This is the language we want to employ to discuss  
	what we 
will call general braneworld holography or just `braneworld holography' in \eqref{bwholo}. While the concept of ``cut-off CFT'' is unspecific, 
the modern interpretation as a $T^2$ theory is very explicit (especially in the $D=3$ case which we will discuss at length) and in some cases allows  
to make non-perturbative statements.

	\subsection{Braneworld holography in terms of $T^2$-deformed CFT} \label{subsect1-2}
	
	We specify first to $D=3$ bulk  
	gravity. In \cite{McGough:2016lol}, it was shown that pulling the CFT into the bulk corresponds to $T\bar T$-deforming it, precisely in the sense that Dirichlet boundary conditions at a finite distance into the bulk give rise to a gravitational path integral that is dual to a $T\bar T$-deformed CFT living on the induced metric 
	\ali{
		Z_{grav}(g) &= Z_{T\bar T}(g) \qquad \text{(cut-off holo for $D=3$)}.  \label{VMM}
	} 
	This is sometimes referred to as cut-off holography. It was  
	conjectured in \cite{McGough:2016lol} for the pure gravity case, which we will mostly be concerned with, and later extended to include bulk matter \cite{Guica:2019nzm,Hartman:2018tkw,WallAraujo-Regado:2022gvw}.  
	The $T\bar T$ theory \cite{Zamolodchikov:2004ce,Smirnov:2016lqw,Cavaglia:2016oda} is obtained from a particular irrelevant deformation of the CFT that is constructed out of stress tensor components in such a way that the initial deformation is the product of the holomorphic and anti-holomorphic stress tensors, hence the name $T\bar T$. The deformation is defined by the flow of the action with respect to the deformation parameter $t$  
	\ali{
		\frac{d}{d t} S^{(t)} 
		_{T\bar T} = \frac{1}{4\pi} \int d^2 x \, \sqrt{-g} \,\left( T_{\mn} T^{\mn} - (T^\mu_{\mu})^2 \right), \qquad S^{(0)}_{T\bar T} = S_{CFT}  \label{TTflow}
	}
	in terms of the stress tensor of the deformed theory $T_{\mu\nu}=\frac{4\pi}{\sqrt{-g}}\frac{\delta S^{(t)}_{T\bar{T}}}{\delta g^{\mu\nu}}$. The duality involves the mapping $c = 12\pi l/\kappa_3$ and $t = -\kappa_3 l/(4\pi)$ between boundary and bulk theory parameters. The $T\bar T$ theory lives on $g$, the fixed induced metric\footnote{In an alternative yet equivalent interpretation, it lives on a rescaled metric and $t$ depends explicitly on the radial location of the boundary.}. 
	For the purposes of this paper, we will not need more info on the $T\bar T$ theory itself, but will just make use  
	of the cut-off holography dictionary \eqref{VMM}.  
	
	One thing to stress here is that the duality \eqref{VMM} is for $Z_{grav}(g)$ given in \eqref{Zgravofg}, with $S_{ct}$ the Balasubramanian-Kraus counterterms, as discussed in more detail below Eq.~\eqref{Sct}. They depend on the induced metric $g$ at $r=\bar r$, while not having explicit dependence on the value of $\bar r$.  
	That is, the counterterms are defined through \eqref{Sct} from the on-shell action contributions that diverge for $\bar r \ra \epsilon$ as $\epsilon \ra 0$ (modulo the $\log \epsilon$ Weyl anomaly term), but are then rewritten in terms of $g$ and are actually finite for finite $\bar r$. 
	The addition of $S_{ct}$ to describe AdS$_{d+1}$-gravity in a `finite box' is ambiguous because of this, but it is the prescription of \cite{McGough:2016lol} that leads to a consistent dictionary \eqref{VMM} with  
	$T\bar T$. 
		Namely, the main checks on the dictionary performed in \cite{McGough:2016lol} such as the map of the radial Wheeler-DeWitt flow for $Z_{grav}$ to a $T\bar T$ flow equation, and the matching of the energy spectrum of the $Z_{grav}(g)$ theory to that of $T\bar T$, require $Z_{grav}(g)$ to be defined as in \eqref{Zgravofg}, with the usual counterterms $S_{ct}$ of \eqref{Zgrav}. 
		This prescription also guarantees a well-defined undeformed limit.   
	To paraphrase, even though the boundary is at a finite distance into the bulk, the counterterms that are added to the gravitational action are the same (as a functional of $g$)
	as you would add in the case of an asymptotic boundary.
	
	$T\bar T$ theory is a $(d=2)$-dimensional theory and is best understood in that case. One can however consider $Z_{grav}(g)$ in higher dimensions $D>3$, 
	and give a name to the corresponding dual theory. That is done in \cite{Hartman:2018tkw,Taylor:2018xcy}, who propose the dictionary  
	\ali{
		Z_{grav}(g) &= Z_{T^2}(g) \qquad \text{(cut-off holo for $D>3$)} .  \label{HK}
	}
	The dual deformed theory is called $T^2$-deformed CFT, with deformation operator 
	quadratic in the stress tensor. It is the 
	modern and improved version of \eqref{AdScutoff}.\footnote{We note that in the extension to higher dimensions issues arise both on the bulk and boundary side of the duality, such as well-definedness of the Dirichlet boundary conditions \cite{Witten:2022xxp} and existence of the $T^2$ operator (although the 
	factorization property is protected by large $c$), which are not well-understood yet.}  
	
	Now we have all the ingredients, using in particular \eqref{Ztot}, \eqref{Zbw}, \eqref{ZtottoZgrav}-\eqref{Sbwgravdef} and \eqref{HK}, 
	to interpret braneworld for  
	general branes (whether in a FG expansion near the boundary or at any finite distance into the bulk) as being holographically dual to $T^2$-deformed CFT coupled to effective gravity on the brane 
	\ali{
		Z_{bw} &= \int Dg \, e^{i \left( W_{\text{$T^2$}}[g] + S_{bwgrav}[g]\right)}   
		\qquad \text{(braneworld holo)}. \label{bwholo} 
	}  
	This is the main result of this section.   
	We are using here the general dimensional notation $T^2$, with the understanding that for $D=3$ it refers to the original $T\bar T$. 
	The effective brane gravity $S_{bwgrav} \equiv S_T - S_{ct}$ is simply determined by the difference between brane tension and counterterms, so as\footnote{The expansion for $S_{bwgrav}$ below should be truncated before divergences arise, depending on $d$.}   
	\ali{
		S_{bwgrav} &= \frac{1}{\kappa_{d+1}} \int_{\p \mathcal M} d^d x \sqrt{-g} \Bigg[(d-1) - T + \frac{1}{2(d-2)}R  \nonumber \\ 
		&\qquad  + \frac{1}{2(d-4)(d-2)^2}\left(R_\mn R^\mn - \frac{d}{4(d-1)}R^2\right) + \cdots \Bigg]. \label{Sbwgravd}}
Here and in the rest of the paper we set $l=1$. 
	This 
	identification of the effective brane theory can also be seen nicely at the level of the EOM by rearranging the Neumann boundary condition or $\delta g$ EOM \eqref{Neumann}, 
	making use of the notation introduced in \eqref{TBY}. With $-\frac{1}{\kappa_D}(K^\mn - K g^\mn)$ written as $T^\mn_{BY \text{w/o ct}}$ or $T^\mn_{BY} - T^\mn_{BY,\text{ ct contrib}}$, and $-\frac{1}{\kappa_D} T g^\mn$ as the tension contribution $T^\mn_{BY,\text{ $T$ contrib}}$ to the Brown-York stress tensor, the Neumann condition 
	\ali{
		K^\mn - K g^\mn + T g^\mn = 0 
	}
	or 
	\ali{
		T^\mn_{BY} - T^\mn_{BY,\text{ ct contrib}} + T^\mn_{BY,\text{ $T$ contrib}} = 0 
	}
	takes the form of an effective gravitational EOM 
	\ali{
		T^\mn_{BY} - \frac{1}{\kappa_d} \mathcal G^\mn
		- \frac{1}{\kappa_d} \Lambda_{d} \, g^\mn = 0    \label{NBCgEOM}
	}
	with the counterterm to tension difference determining the $d$-dimensional Einstein tensor $\mathcal G_\mn$ as well as 
	cosmological constant $\Lambda_d$ term.  
	This identification  
	$T^\mn_{BY,\text{ ct contrib}} -  T^\mn_{BY,\text{ $T$ contrib}} \equiv \frac{1}{\kappa_d} \mathcal G^\mn + \frac{1}{\kappa_d} \Lambda_{d} \, g^\mn$  
	is valid in $D=4,5$ where the effective theory $S_{bwgrav}$ takes a $d$-dimensional Einstein gravity form, and is replaced by the higher-derivative gravity equivalent in higher dimensions, depending on the form of $S_{ct}$ in \eqref{Sbwgravd}.  
	The effective cosmological constant $\Lambda_d$ will  
	have a contribution proportional to the tension $T$ plus contributions from the counterterm. 
	The Brown-York stress tensor by \eqref{HK} is dual to the expectation value of the $T^2$ stress tensor, so that in boundary notation the brane theory EOM for $d=3,4$ are given by
	\ali{
		\mathcal G^\mn 
		+ \Lambda_{d} \, g^\mn = \kappa_d \vev{T^\mn}   . \label{bweinsteineom}
	}

	The curvature terms that appear in $S_{ct}$ are familiar both from braneworld holography ($S_{bwgrav}$) as well as $T^2$ holography ($W_{T^2}$) \cite{Hartman:2018tkw,Taylor:2018xcy}, precisely because one rewrites the braneworld theory $S_{tot}$ into $W_{T^2} + S_{bwgrav}$ by adding and subtracting the counterterms $S_{ct}$, 
	\ali{
		S_{tot} = (S_{EH}+S_{GH}+S_{ct}) - (S_{ct}-S_T) = W_{T^2} + S_{bwgrav}    
	}
	with the curvature terms of $S_{ct}$ thus contained both in $W_{T^2}$ 
	and $S_{bwgrav}$.  
Applying the general dictionary \eqref{bwholo}, the statement becomes that AdS$_{D}$-gravity bounded by an EOW brane is holographically dual to the $d$-dimensional \emph{$T^2$-deformed CFT} coupled to the effective braneworld gravity $S_{bwgrav}$ given in \eqref{Sbwgravd}.  
	In particular, for $D>3$ the braneworld gravity is governed by the action 
\ali{
	S_{bwgrav}&=\frac{1}{2 \kappa_d}\int d^dx \sqrt{-g}(R-2 \Lambda_d)+\dots 
	\\
	\text{with} \; \; \; \kappa_d=&\kappa_{d+1}(d-2), \; \; \; \Lambda_d=(d-2)(1-d+T) \label{SBWgravdparameters}}
with the dots denoting higher curvature corrections appearing in $D>5$. For example, in the $D=5$ case discussed in \cite{Gubser:1999vj}, the braneworld gravity in \eqref{Sbwgravd} is identified with $S_{bwgrav} = \frac{1}{2 \kappa_4} \int \sqrt{-g} (R - 2 \Lambda_{4})$, with the bulk and braneworld parameters related as $\kappa_4 = 2 \kappa_5$ and $\Lambda_{4} = 2T-6$\footnote{Expanding around a flat saddle $\hat{g}$ fixes the tension to $T=d-1$, such that $\Lambda_d=0$.}.  
In $D=3$, things are a bit more complicated as we go on to discuss next.

\section{$3D$ braneworld holography: Holographic Weyl anomaly set free} \label{sectWeyl} 
	
	In the case of a 3-dimensional bulk, the counterterm action is just an area term, which we can read off from \eqref{Sbwgravd} to give  
	\ali{
		S_{bwgrav} = \frac{1}{\kappa_3} \int d^2x \sqrt{-g} (1-T) .  \label{Sbw3D}
	}
	The effective gravity is 
	reduced to a pure cosmological constant term $S_{bwgrav} = -\frac{1}{\kappa_2} \int \sqrt{-g} \Lambda_2$ with $\kappa_2=2\kappa_3$ and the cosmological constant $\Lambda_2=2(T-1)$ determined by the tension of $S_T$ shifted by a number coming from $S_{ct}$.  
	Before discussing $3D/2d$  
	general braneworld in section \ref{sectTT}, 
	let us first think about the asymptotic brane case. 
	There are no kinetic terms for the metric $g_\mn$ in $S_{bwgrav}$ in \eqref{Sbw3D}. Instead, one can make use of the integrated Weyl anomaly to extract an action from $Z_{CFT}(g_{(0)})$ that contains kinetic terms for the conformal factor 
	of the metric. This will be the Liouville action, 
	taking the role analogous to that of the Einstein-Hilbert action for $d=3,4$ in the braneworld theory, with central charge $c$ in the role of $1/\kappa_2$.   
	While this is the usual interpretation of $3D/2d$ (asymptotic) braneworld holography \cite{Compere:2008us,Suzuki:2022xwv},  
	it is hard to find a detailed discussion. For us, it provides the starting point for the 
	extension to a non-asymptotic braneworld discussion in section \ref{sectTT}. Therefore, we will spend a whole section on the interpretation of $3D/2d$ braneworld holography in terms of Liouville theory. This requires a revisiting of the holographic Weyl anomaly, particularly the integrated Weyl anomaly. Subsection \ref{subs2d} and \ref{subs3d} will provide the ingredients for the interpretation of the  $Z_{bw}$ theory at hand in subsection \ref{subsholo}. 
	
	As the rest of the paper is focused on the $D=3$ set-up, we will from now on 
	use the notation $\kappa$ for $\kappa_3$ and $\lambda$ for (minus) $\Lambda_2$.

	\subsection{Liouville description of Weyl anomaly in $2d$ CFT} \label{subs2d} 
	
We start with a CFT section that reviews the integrated Weyl anomaly, to set our notation. We introduce a `triangle representation' for a conformal transformation, whose bulk interpretation will prove useful for the strategy of the bulk calculations, and discuss the role of the Liouville stress tensor.

	We consider a $2d$ CFT with path integral $Z_{CFT}(g)$ and central charge $c$. 
	It depends on the source field $g$ being the background metric. 
	Locally, any $2d$ metric is conformally flat, so we write\footnote{The notation below is restricted to section \ref{subs2d}, which is purely about $2d$ CFT. In the remainder of the paper, which focuses on holographic CFT's, we will instead denote the background geometry of the CFT as $g_{(0)}$ (the conformal boundary metric), with Weyl mode 
			$\phi$, i.e.~$g_{(0)}=e^{\phi}\hat{g}_{(0)}$. These will be distinguished from the background geometry of the holographic $T\bar{T}$ theory, denoted as $g$ (the induced metric on the boundary), with 
			Weyl mode $\sigma$, i.e.~$g=e^{\sigma}\hat{g}$.}  
	\ali{
		g_\mn = e^\phi \hat g_\mn  \label{2Dmetric}
	}
	with fixed reference metric $\hat g$ of the form $df d\bar f$.  
	The Weyl anomaly of the CFT can be stated
	as the fact that 
	$Z_{CFT}$ has to satisfy the conformal Ward identity 
	\ali{
		\frac{1}{\sqrt{-g}} \frac{\delta}{\delta \phi} Z_{CFT}(g) = i \frac{c}{48\pi} (R + \lambda) \,  Z_{CFT}(g)   \label{Wardid}
	} 
	where we have included the possibility of a constant $\lambda$ in the Weyl anomaly \cite{Fabbri:2005mw}. 
	This identity can be integrated to what is then called the integrated Weyl anomaly  \cite{ZZ_Liouville_gravity,Erbin2015_2dGravityLiouville}
	\ali{
		Z_{CFT}(g) = Z_{CFT}(\hat g) \, e^{i S_L[\phi; \hat g]}   
		\label{integratedWeyl}
	}
	with $S_L$ the Liouville action for Liouville field $\phi$ 
	\ali{
		S_L &= -\frac{c_L}{48 \pi} \int d^2 x \sqrt{-\hat g} \left( \phi \, \hat R + \frac{1}{2} \hat g^{\mn} \p_\mu \phi \p_\nu \phi + \lambda \, e^{\phi} \right). \label{SLWeyl}
	} 
	The Liouville central charge $c_L$ is given by $c_L = -c$ in terms of the CFT central charge. For positive $c$, the kinetic term has the `wrong' sign, therefore the Liouville theory is timelike. 
	In terms of the stress tensor $\langle T_{\mu\nu} \rangle = \frac{4\pi}{\sqrt{g}} \frac{\delta W_{CFT}}{\delta g^\mn}$, 
	the Weyl anomaly \eqref{Wardid} 
	expresses the non-vanishing of the trace 
	\ali{
		\vev{T_\mu^\mu} = -\frac{c}{12} (R + \lambda)  \label{traceT} 
	} 
	which can be traced back to the transformation behavior of the CFT stress tensor containing an anomalous Schwarzian contribution proportional to $c$.   
	For later reference, the stress tensor associated with the Liouville action 
	$t_\mn^L = \frac{4\pi}{\sqrt{\hat g}} \frac{\delta S_L}{\delta \hat g^\mn}$  
	is given by 
	\ali{
		t_\mn^L = 
		\frac{c_L}{24} \left(-\p_\mu \phi \p_\nu \phi + \hat g_\mn \left( \frac{1}{2} \hat g^\ab \p_\al \phi \p_\beta \phi + \lambda e^\phi \right) + 2 (\hat \nabla_\mu \p_\nu \phi - \hat g_\mn \hat \Box \phi ) \right). \label{tL} 
	}
	
	Let us use the language of \cite{Fulton:1962bu} for thinking about a (active) conformal transformation as the combination of a point transformation followed by a passive coordinate transformation. 
	In the $2d$ boundary manifold with metric $\hat g(x) dx^2$, a point transformation  
	$x \ra \tilde x$ is called 
	conformal when the metric 
	evaluated in the new point is proportional to the metric in the point $x$, i.e.~$\hat g(\tilde x) d\tilde x^2 = \Omega(x) \hat g(x) dx^2$.      
	It is followed by a change of frame with the property $x' \circ \tilde x (x) = x$, such that the final metric $g(x) dx^2$ 
	is related to the original metric $\hat g(x) dx^2$ by an active conformal transformation, 
	$g(x) = \Omega(x) \hat g(x)$.  
	This is summarized in what we will refer back to as the `boundary triangle' notation  
\begin{equation}
	\begin{tikzcd}[row sep=3cm, column sep=4cm]
		\begin{array}{c}  
			x = (f, \bar f) \\[8pt] 
			g(x)\, dx^2 \;=\; \Omega \, df \, d\bar f  
		\end{array}
		&
		\begin{array}{c}  
			\tilde{x} = (z,\bar z) \\[8pt]  
			\hat{g}(\tilde{x})\, d\tilde{x}^2 \;=\; dz \, d\bar z 	
		\end{array}
		\arrow[l, "\;\; passive \;\;"]
		\\
		&
		\begin{array}{c}  
			x = (f,\bar f) \\[8pt] 
			\hat{g}(x)\, dx^2 \;=\; df \, d\bar f 
		\end{array}
		\arrow[u, "\;\; point \;\;"] 
		\arrow[ul, "\;\; conformal \;\;"]
	\end{tikzcd}
	\label{bdytriangle1}
\end{equation}
	The (active) conformal transformation $x \ra \tilde x$ or 
	$f \ra z$ connects the lower right and upper left corner, and connects conformally related metrics. 
	In the top line, however, the line elements are related by a passive coordinate transformation step and thus equal. 
	We have included  
	in the schematic representation $2d$ notation 
	and a starting metric $ds^2 = df d\bar f$,   
	for which it is clear that 
	\ali{
		\Omega = \frac{\p z}{\p f} \frac{\p \bar z}{\p \bar f} \, . \label{Omega}  
	} 
	
	For a Virasoro primary operator $\hat {\mathcal O}$ of dimension $h = \bar h$, we know from its defining transformation behavior that $\vev{\mathcal O(f,\bar f)}_{\Omega \, df d\bar f} = \Omega^{h} \vev{\hat {\mathcal O}(z,\bar z)}_{dz d\bar z}$ in the top line of the triangle. Conformal invariance  $\vev{\mathcal O(f,\bar f)}_{\Omega \, df d\bar f} = \vev{\hat{\mathcal O}(f,\bar f)}_{df d\bar f}$ under the $f \ra z$ transformation (along the diagonal in the triangle notation) 
	is then expressed as 
	\ali{
		\vev{\hat{\mathcal O}(f,\bar f)}_{df d\bar f} = \Omega^{h} \vev{\hat{\mathcal O}(z,\bar z)}_{dz d\bar z} . \label{correl} 
	} 
	We have explicitly included as a subscript the background metrics in which these correlation functions are taken, having in mind really correlators $\vev{\mathcal O \cdots \mathcal O}$ with operators at different locations  
	certain distances apart. \eqref{correl} in particular relates correlators in conformally related metrics. 
	For the stress tensor, 
	\ali{
		\vev{{\hat T_{ff}}(f)}_{df d\bar f} = \left|\frac{\p z}{\p f}\right|^{2} \vev{{\hat T_{zz}}(z)}_{dz d\bar z} + \frac{c}{12} \{z,f\} \label{Ttransf}  
	} 
	with $\{z,f\} \equiv \frac{z'''}{z'} - \frac{3}{2} \frac{z''^2}{z'^2}$ the Schwarzian derivative.   
	(A typical example is the plane to cylinder transformation $f=e^z \ra z$, for which $\vev{\hat T(f)}_{df d\bar f} = 0$ and $\vev{\hat T(z)}_{dz d\bar z} 
	= \frac{c}{12} \{f,z\} = -c/24$. 
	More generally the stress tensor expectation values 
	for a given geometry can be obtained from the Schwarzian of the uniformizing coordinate with respect to the geometry coordinate.)  
	One can think of the anomalous Schwarzian term  
	as measuring the failure of conformal invariance of stress tensor correlators (along the diagonal in the triangle) when the stress tensor is assumed to transform as a regular tensor. Alternatively, \eqref{Ttransf} can be read as the required anomalous transformation behavior of the stress tensor for the conformal invariance to hold.   
	
	The Weyl factor $\Omega$ can be written as $e^\phi$, as in \eqref{2Dmetric}. For $\Omega$ given in \eqref{Omega}, the field $\phi(f,\bar f)$ is 
	\ali{
		\phi = \log \left( \frac{\p z}{\p f} \frac{\p \bar z}{\p \bar f} \right).  \label{phisol}
	} 
	In this notation,  
	the anomalous term in \eqref{Ttransf} is a Liouville stress tensor \eqref{tL}, 
	\ali{
		\frac{c}{12} \{ z,f \} = -t_{ff}^L[\phi] 
	}
	with 
	\ali{
		t_{ff}^L
		&= \frac{c}{24} \left( (\p_f \phi)(\p_f \phi) 
		- 2 \p_f^2 \phi \right) . 
	}  
	Then the diagonal arrow of conformal invariance expresses 
	\ali{
		\vev{T_{ff}}_{e^\phi df d\bar f} &= \vev{\hat T_{ff}}_{df d\bar f} \, + \,  t_{ff}^L[\phi]  \label{LBmROtensor}   
	}
	if the first term on the RHS in \eqref{Ttransf} is rewritten as $\vev{T_{ff}}_{e^\phi df d\bar f}$, i.e.~in the interpretation with assumed regular tensor transformation behavior (which will be the natural interpretation from a bulk perspective later).  
	
	\eqref{LBmROtensor} is indeed consistent with the separation of a Liouville theory  
	from the CFT path integral $Z_{CFT}$ in the integrated Weyl anomaly \eqref{integratedWeyl}. To see this more explicitly, we write out the variation $\delta \log Z_{CFT}(g)$, equal to $\frac{1}{4\pi i}  \int d^2 x \sqrt{-g} \vev{T^\mn} \delta g_\mn$, by making use of \eqref{integratedWeyl}, $\delta \log Z_{CFT}(\hat g) = \frac{1}{4\pi i}  \int d^2 x \sqrt{-\hat g} \vev{\hat T^\mn} \delta \hat g_\mn$ and writing out the full metric variation as 
	\ali{
		\delta g_\mn = e^\phi \delta \hat g_\mn +  g_\mn \delta \phi. 
	}   
	This gives\footnote{Note that $\vev{T_\mn} = t^L_\mn + \vev{\hat T_\mn}$ but 
		$\vev{T^\mn} = e^{-2\phi}(t_L^\mn + \vev{\hat T^\mn})$.} 
	\ali{
		\delta W_{CFT}[g] 
		&=- \frac{1}{4\pi}  \int d^2 x \sqrt{-\hat g}  \left(  t_L^\mn + \vev{\hat T^\mn} \right) \delta \hat g_\mn  - \frac{c_L}{48\pi} \int d^2 x \sqrt{-\hat g} \, (\mathcal L \text{ EOM}) \, \delta \phi \label{deltalogZCFT}
	}
	where the Liouville equation of motion in the last term is $\sqrt{-\hat g} \, (\mathcal L \text{ EOM}) = \sqrt{-g}(R + \lambda)$, representing the trace part of the stress tensor \eqref{traceT}, or in terms of the hatted variables,  $(\mathcal L \text{ EOM}) = \hat R + \lambda e^\phi  - \hat g^\mn \hat \nabla_\mu \p_\nu \phi$.  
	For our flat $\hat g$ and $\lambda = 0$, it expresses the vanishing of $\p_f \p_{\bar f} \phi$, which is satisfied by our Weyl mode \eqref{phisol}.  
	The split of the $\delta \hat g$ contribution in a Liouville stress tensor and a stress tensor for $Z_{CFT}(\hat g)$ is then indeed consistent with \eqref{LBmROtensor}.

	\subsection{Holographic integrated Weyl anomaly} \label{subs3d}

	We now derive the Liouville action $S_L$ in \eqref{integratedWeyl} from a bulk perspective. 
	
	\paragraph{Strategy}
	
	Our starting point is an asymptotically AdS$_3$ metric in Fefferman-Graham coordinates 
	\ali{
		\hat G(X)dX^2 = d\rho^2 + e^{2\rho} \left( \hat g_{(0)}(x) + e^{-2\rho} \hat g_{(2)}(x) + \cdots \right) dx^2    \label{FG1}
	} 
	with $\hat g_{(0)}$ given as boundary condition. We will denote it 
	$\widehat{FG}(\rho,x^\mu)$. 
	Mimicking the boundary procedure for performing a conformal transformation, we can push the points $X = (\rho, x^\mu)$ to $X' = (\rho', x'^\mu)$. In this intermediate step we obtain a metric 
	\ali{
		\hat G(X')dX'^2 = d\rho'^2 + e^{2\rho'} \left( \hat g_{(0)}(x') + e^{-2\rho'} \hat g_{(2)}(x') + \cdots \right) dx'^2    \label{FGp}  
	}
	which we will refer to as $\widehat{FG}(\rho',x'^\mu)$. 
	Then we  
	perform a coordinate transformation 
	$X'_{(\phi)}(X)$, labeled by a function $\phi$, which is designed to take us 
	to \emph{another} Fefferman-Graham (i.e.~asymptotically AdS$_3$) metric expressed in the original coordinates,  
	\ali{
		G(X)dX^2 = d\rho^2 + e^{2\rho} \left( g_{(0)}(x) + e^{-2\rho} g_{(2)}(x) + \cdots \right) dx^2.     \label{FG2}
	}  
	This represents a 
	different Fefferman-Graham line element $FG(\rho,x^\mu)$, with accordingly different  
	metric fields $g_{(i)}(x)$ in the expansion. The last step uses $G(X)dX^2 = \hat G(X')dX'^2$. 
	
	The coordinate transformation $X'_{(\phi)}(X)$ required to go from one asymptotically AdS to another asymptotically AdS metric is a Brown-Henneaux (BrH) coordinate transformation \cite{Brown:1986nw}. It can be constructed order by order in $e^{-2\rho}$ as follows. 
	One starts with a coordinate transformation ansatz in the form of an asymptotic expansion   
	and parametrized by a field $\phi(x)$, 
	\ali{
		\begin{split}
			\rho'(\rho,x) &= \rho + \frac{1}{2}\phi(x) + \sum_{j=1} e^{-2\rho j} a_\rho^{(2j)}(x) \\ 
			x'^\mu(\rho,x) &= x^\mu + \sum_{j=1} e^{-2\rho j} a_x^{(2j) \mu}(x).
		\end{split} \label{BrHexp}
	}
	The leading behavior has been separated out in the notation and could alternatively be written as $a_\rho^{(0)} = \phi/2$ and $a_x^{(0)} = 0$. 
	The unknown functions are determined at each order by imposing  $G_{\rho\mu}=0$ and $G_{\rho\rho}=1$, giving rise at first subleading order to 
	\ali{
		a_\rho^{(2)} = \frac{1}{16} e^{-\phi} \hat g^{\mn}_{(0)} \p_\mu \phi \p_\nu \phi, \qquad a_x^{(2)\mu} = \frac{1}{4} e^{-\phi} \hat g^{\mn}_{(0)} \p_\nu \phi. 
	}
	The expansion is in $e^{-2\rho}$ and $\phi$ is finite so these are order-by-order finite Brown-Henneaux diffeomorphisms. 
	The infinitesimal ones were discussed e.g.~in \cite{Imbimbo:1999bj} and the finite ones in the form \eqref{BrHexp} in \cite{Skenderis:2000in}. 
	
	The resulting $G_\mn$ in \eqref{FG2} to first order 
	contain  
	\ali{
		g^{(0)}_{\mn} &= e^{\phi} \hat g^{(0)}_{\mn}  \label{g0phi}
	}
	and 
	\ali{
		g^{(2)}_{\mn} &= \hat g^{(2)}_{\mn} - \frac{1}{4} \p_\mu \phi \p_\nu \phi + \frac{1}{2} \hat \nabla^{(0)}_\mu \p_\nu \phi + \frac{1}{8} \hat g^{(0)}_{\mn} \hat g^{\alpha\sigma}_{(0)} \p_\alpha \phi \p_\sigma \phi. \label{g2phi} 
	}  
	The Brown-Henneaux diffs $X \ra X'_{(\phi)}(X)$  
	thus connect the FG metric $\widehat{FG}(\rho,x^\mu)$ in \eqref{FG1}, with conformal boundary $\hat g_{(0)}$, to the FG metric $FG(\rho,x^\mu)$ in \eqref{FG2} 
	with conformal boundary $e^{\phi} \hat g_\mn^{(0)}$.   
	As such, we have completed a bulk extension of the boundary triangle notation of 
	a conformal transformation in \eqref{bdytriangle1}, relating boundary metrics that differ by a Weyl factor $e^{\phi}$.  
This is summarized as the `bulk triangle' 
     \begin{equation}
     	\begin{tikzcd}[row sep=3cm, column sep=4cm]     	
     		\begin{array}{c}  
     			X = (\rho, f, \bar f) \\[8pt] 
     			G(X)\, dX^2 \;=\; F G(\rho, x^\mu)   
     		\end{array}
     		&     		
     		\begin{array}{c}  
     			X' = (\rho', y, \bar y) \\[8pt]  
     			\hat{G}(X')\, dX'^2 \;=\; \widehat{FG}(\rho', x'^\mu) 	
     		\end{array}
     		\arrow[l, "\;\; passive \;\;"]
     		\\     		
     		&
     		\begin{array}{c}  
     			X = (\rho, f, \bar f) \\[8pt] 
     			\hat{G}(X)\, dX^2 \;=\; \widehat{FG}(\rho, x^\mu)  
     		\end{array}
     		\arrow[u, "\;\; point \;\;"] 
     		\arrow[ul, "\;\; BrH \,\, di\mathit{ff} \;\;"]
     	\end{tikzcd}
     	\label{bulktriangle}
     \end{equation}
where we have included boundary notation for the longitudinal coordinates\footnote{By longitudinal coordinates, we mean coordinates along the boundary, perpendicular to the bulk radial direction.}.  
As an example, we will later specify to a Poincar\'e geometry for $\widehat{FG}(\rho,x^\mu)$ with conformal boundary $df d\bar f$, but the derivation holds for general $\widehat{FG}$.

	What we are left to do is compare the 
	on-shell gravitational actions along the diagonal of the triangle  
	\ali{  
		S_{grav}[FG(\rho,x^\mu)] 
		- S_{grav}[\widehat{FG}(\rho,x^\mu)] \label{FGmhatFG}
	} 
	in order to compare the corresponding large $c$ CFT theories, i.e.~holographically calculate 
	\ali{
		W_{CFT}(g_{(0)}) - W_{CFT}(\hat g_{(0)}) ,  \label{WCFTsubtraction} 
	}
	with the notation in \eqref{FGmhatFG} 
	referring to evaluation on the respective bulk geometries.   
	This is expected to produce precisely the Liouville action $S_L$ of \eqref{integratedWeyl} describing the Weyl anomaly of the dual CFT. 
	
	Of course, the notation in \eqref{FGmhatFG} is schematic and incomplete. 
	We need to specify the integration limits of the action integrals evaluated on $\hat G$ and $G$, such that the correct boundary metrics $\hat g_{(0)}$ and $g_{(0)}$ are compared.   
	This is the subtle part of the calculation. To proceed, let us first  
	more carefully understand how the boundary triangle \eqref{bdytriangle1} fits in the bulk triangle \eqref{bulktriangle}.

\begin{figure}
    	\centering
    	\includegraphics[scale=0.4]{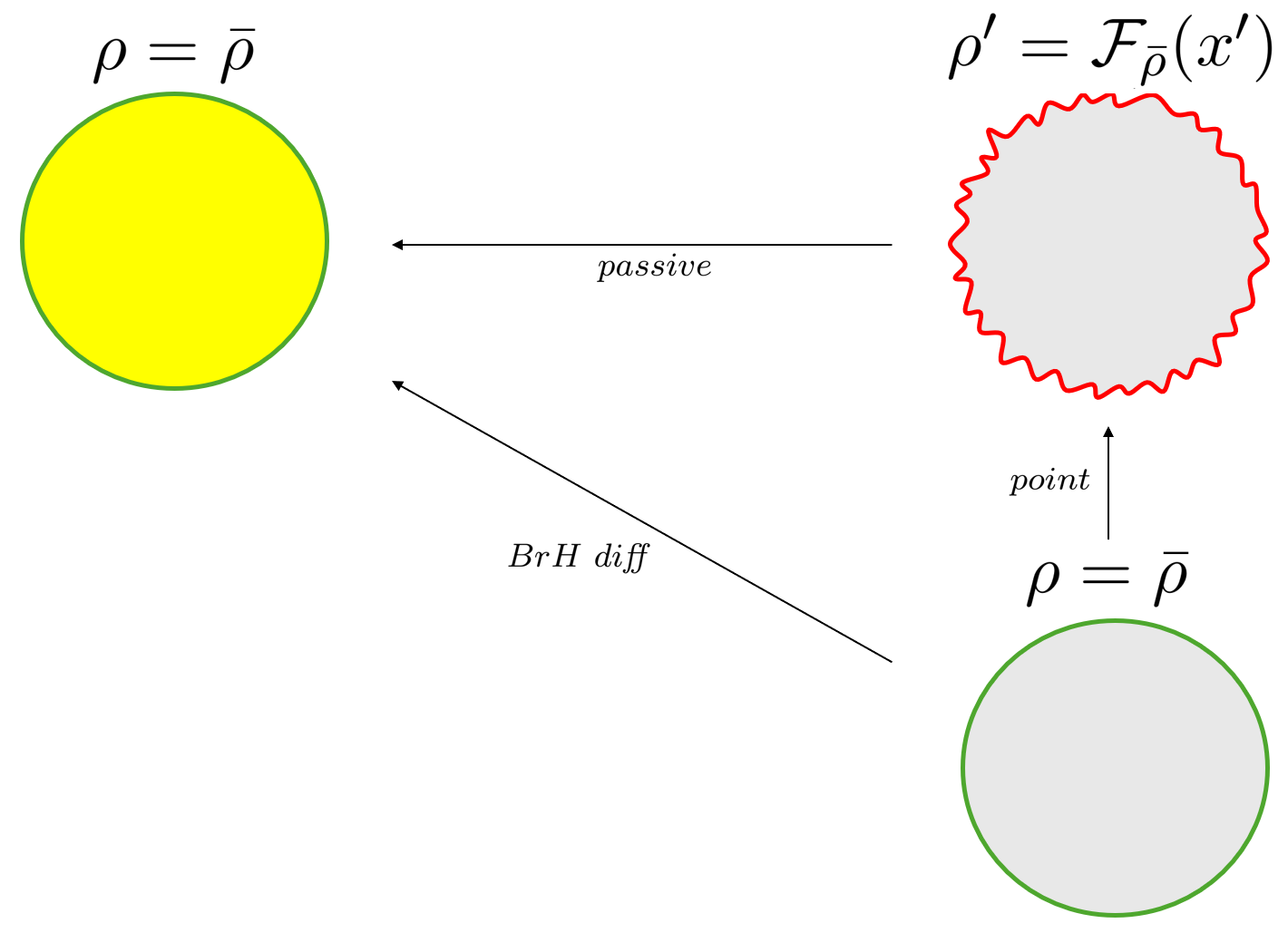}
    	\caption{
Pictorial `bulk triangle' representation \eqref{bulktriangle} of a Brown-Henneaux diffeomorphism taking us from a bulk metric $\hat G(X) dX^2$ in the lower right corner to a bulk metric $G(X) dX^2$ in the upper left corner along the diagonal arrow, via a point transformation (vertical arrow) to $\hat G(X') dX'^2$ and passive coordinate transformation (left-pointing arrow). The grey color indicates the metric $\hat G$ and yellow the metric $G$. With the boundaries included as sketched, the three pictures refer to the bulk metrics on which to evaluate the gravitational action in order to holographically compute the Liouville action $S_L$ from \eqref{WCFTsubtraction} for $\bar \rho \ra \infty$, or its higher-derivative extension $S_{\tilde L}$ in \eqref{Wttbarsubtraction} for finite $\bar \rho$. 
    	} 
    	\label{trianglecirclesfigure}
    \end{figure}

	We consider constant $\rho$ boundaries at $\rho = \bar \rho$ of $\widehat{FG}(\rho,x^\mu)$ and $FG(\rho,x^\mu)$ in the bulk triangle representation of the Brown-Henneaux diffeomorphism, with $\bar \rho \ra \infty$ induced metrics respectively equal to $e^{2\bar \rho} \hat g_{(0)}(x)dx^2$ and $e^{2\bar \rho} e^{\phi(x)} \hat g_{(0)}(x)dx^2$.	 
It is clear that 
these respectively reach the Dirichlet fixed conformal boundary metrics $\hat g_{(0)}(x)dx^2$ and $e^{\phi(x)} \hat g_{(0)}(x)dx^2$ in the corresponding boundary triangle representation of the conformal transformation. 
To find the location of the co-dimension one hypersurface in  $\widehat{FG}(\rho', x'^\mu)$ 
with 
induced metric 
given asymptotically by $e^{2\bar \rho} e^{\phi(x)} \hat g_{(0)}(x) dx^2$,  
one must simply follow the Brown-Henneaux coordinate transformation \eqref{BrHexp} back from $\rho = \bar \rho$ in $FG(\rho, x^\mu)$: 
\ali{
	&\rho' = \rho'\left(\bar \rho, x(\bar \rho, x')\right) = \mathcal F_{\bar \rho}(x'), \label{curlycutoff}  \\
	&\text{with }\mathcal F_{\bar \rho}(x') = \bar \rho + \frac{1}{2} \phi(x') + \mathcal O(e^{-2\bar \rho}).  
}	
Here we needed the inverse Brown-Henneaux transformation of the longitudinal coordinates, which can be obtained from \eqref{BrHexp} order by order in the asymptotic expansion.   
The function $\mathcal F_{\bar \rho}$ specifies the non-constant value of $\rho'$ at which the induced metric reaches (for $\bar \rho \ra \infty$) 
the conformal boundary 
$\hat g_{(0)}(\tilde x) d\tilde x^2$ in the top right corner of the boundary triangle. 
The location depends on the longitudinal coordinate, describing a `curly boundary' \footnote{
	Clearly, it is not so that the constant $\rho'$ boundary of $\widehat{FG}(\rho',x'^\mu)$ is given by the corresponding top right corner $\hat g_{(0)}(\tilde x) d\tilde x^2$ in the boundary triangle. This follows simply from the equalities $\hat g_{(0)}(\tilde x) d\tilde x^2 = e^{\phi(x)} \hat g_{(0)}(x) dx^2$ and $G(X)dX^2 = \hat G(X')dX'^2$ in the top lines of both triangles, combined with the fact that the bulk radial coordinates differ, $\rho \neq \rho'$ (and thus generally $d\rho^2 \neq d\rho'^2$). 	
}, and depends parametrically on the constant $\bar \rho$ that `remembers' the original location in $FG(\rho,x^\mu)$. 
This is depicted in Fig.~\ref{trianglecirclesfigure} \footnote{ 
	Essentially, \eqref{curlycutoff} is a simple statement about the location of the boundary changing under the non-trivial mapping. 
	We expanded on it because it justifies the interpretation on p.~25 of the Liouville theory calculated from comparing the actions along the vertical arrow in Fig.~\ref{trianglecirclesfigure} as the Weyl anomaly (which relates the actions along the diagonal arrow in Fig.~\ref{trianglecirclesfigure}), 
	and because it puts the use of the radion $\tilde \phi$ in braneworld in a clear holographic perspective (in section \ref{subsectTTWeyl} and section \ref{sectbw}).}. 
Though we will keep the notation and calculation general, typically we will have in mind a plane to cylinder transformation on the boundary corresponding to a Poincar\'e to BTZ bulk transformation\footnote{ 
	The equivalence between a fixed boundary in BTZ to a curly boundary in Poincar\'e can for example be used to calculate the small-interval holographic entanglement `$\log \sinh$' formula from a Ryu-Takayanagi geodesic in either picture.}.  

	We can now refine the subtraction \eqref{FGmhatFG} to 
	\ali{
		\lim_{\bar \rho \ra \infty} \left( S_{grav}[FG(\rho < \bar \rho,x^\mu)] - S_{grav}[\widehat{FG}(\rho < \bar \rho,x^\mu)] \right) \label{FGmhatFGrefined} 
	}
	containing now notation for the upper integration limit. 
	Particularly, having understood the equivalence between the fixed boundary in $FG$ and the curly boundary in $\widehat{FG}$, 
	we know that 
	\ali{
		S_{grav}[FG(\rho < \bar \rho,x^\mu)] = S_{grav}[\widehat{FG}(\rho'< \mathcal F_{\bar \rho}(x'),x'^\mu)] \label{equalityS} 
	}
	and 
	we could alternatively calculate  
	\ali{ 
		\lim_{\bar \rho \ra \infty} \left( S_{grav}[\widehat{FG}(\rho'< \mathcal F_{\bar \rho}(x'),x'^\mu)] 
		- S_{grav}[\widehat{FG}(\rho < \bar \rho,x^\mu)]  \right)  \label{hatFGmhatFG}
	} 
	to holographically obtain the Liouville action. It will turn out that the latter is in fact the more straightforward way.

	\paragraph{Calculation} 
	
	Since all the relevant bulk geometries that we want to evaluate the 
	action on are of the Fefferman-Graham form, it is most useful to start with a general calculation of $S_{grav}$ evaluated on the general \eqref{FG2}, integrated up to a general boundary location 
	\ali{
		\rho = F(x) = \bar \rho +  
		\tilde \phi(x) 
	}  
	with the property that the function $F(x)$ is of the prescribed order $\mathcal O(\bar \rho)$ form and thus large (as $\bar \rho$ is taken to infinity at the end of the calculation). The newly introduced field $\tilde \phi(x)$ is thus of order unity $\mathcal O(1)$. The parametric dependence on the constant $\bar \rho$ has been left out of the $F(x)$ notation for clarity.   
	This leads to a gravitational action in terms of $g_{(0),(2),...}$ in an $e^{-2F}$ or equivalently $e^{-2\bar\rho}$ expansion, given by
	\ali{
		& S_{grav}[FG(\rho < F(x), x)] \nonumber \\
		&\quad =  \frac{1}{2\kappa}\int d^2x \sqrt{-g_{(0)}}\left(g_{(0)}^{\mn}\partial_\mu F\partial_\nu F - g_{(0)}^{\mn} g^{(2)}_{\mn}\left(1+2 F\right)+\mathcal{O}(e^{-2F})\right)+\cdots . 
		\label{SgravFGgeneral}
	} 
	The derivation is delegated to Appendix \ref{appendix:A}. The dots in the expression \eqref{SgravFGgeneral} refer to contributions from the \emph{lower} bound of integration in $\rho$, which we will discuss shortly. 
	  
	We can now go ahead and apply the general result \eqref{SgravFGgeneral} to evaluate \eqref{FGmhatFGrefined} and \eqref{hatFGmhatFG}. 
	Due to the difference $\sqrt{-g_{(0)}} g_{(0)}^{\mn} g^{(2)}_{\mn} - \sqrt{-\hat g_{(0)}} \hat g_{(0)}^{\mn} \hat g^{(2)}_{\mn}$ being a total derivative,   
	we find that 
	\eqref{FGmhatFGrefined} in fact vanishes, 
	\ali{
		\lim_{\bar \rho \ra \infty} \left( S_{grav}[FG(\rho < \bar \rho,x^\mu)] - S_{grav}[\widehat{FG}(\rho < \bar \rho,x^\mu)] \right) = \Delta(\cdots) \label{Deltalbc}  
	} 
	up to the difference in the lower bound contributions, denoted by $\Delta(\cdots)$ on the right hand side.  
	The 
	expression \eqref{hatFGmhatFG}, however, evaluates to   
	\ali{
		&\lim_{\bar \rho \ra \infty} \left( S_{grav}[\widehat{FG}(\rho'< \mathcal F_{\bar \rho}(x'),x'^\mu)]  
		- S_{grav}[\widehat{FG}(\rho < \bar \rho,x^\mu)] \right)  \nonumber\\ 
		&= \frac{1}{2\kappa} \int d^2 x   \sqrt{-\hat g_{(0)}} \left(\frac{1}{4} \hat g_{(0)}^{\mn}\partial_\mu \phi \partial_\nu \phi - \hat g_{(0)}^{\mn} \hat g^{(2)}_{\mn} \, \phi \right).  \label{Liouholo}
	}
This subtraction refers to the vertical arrow in the triangle \eqref{bulktriangle},  comparing the same bulk metrics in different bulk coordinates, which are dummy variables in the action. 
	Lower bound contributions $(\cdots)$ cancel each other in the difference, because we are  
	comparing the action  
	with integration from the \emph{same} lower bound ($\rho = \rho_+$ in the bottom right and $\rho' = \rho_+$ in the top right, suppressed in the notation), just up to different asymptotic boundary locations. 
	For the case of $\widehat{FG}$ being a Poincar\'e AdS solution, for example, the lower integration bound would be the Poincar\'e horizon $\rho_+ \ra -\infty$.   
	
The derivation of \eqref{SgravFGgeneral} makes use of bulk on-shellness. The starting metric $\widehat{FG}(\rho,x^\mu)$ 
therefore obeys the relation \cite{Skenderis:1999nb}
	\ali{
		\hat g_{(0)}^{\mn} \hat g^{(2)}_{\mn} = -\frac{1}{2} \hat R_{(0)} \label{hatFGonshell}  
	} 
	between $\hat g_{(2)}$ and $\hat g_{(0)}$. 
	The resulting action \eqref{Liouholo} is then indeed the expected Liouville action  
	\eqref{SLWeyl} for the dual CFT on $e^\phi \hat g_{(0)}$ and with Brown-Henneaux central charge $c = 12 \pi l/\kappa$,  
	\ali{
		S_L &= \frac{c}{48 \pi} \int d^2 x \sqrt{-\hat g_{(0)}} \left( \phi \,  \hat R_{(0)} + \frac{1}{2}  \hat g^{\mn}_{(0)} \p_\mu \phi \p_\nu \phi  
		\right). \label{SLholoWeyl}
	} 
	There is no Liouville cosmological constant contribution of the form $\lambda e^\phi$ 
	for describing the Weyl anomaly $\vev{T_\mu^\mu} = -(c/12) R_{(0)}$ (with no $\lambda$) in \eqref{traceT}. 
	This completes the holographic derivation of the integrated Weyl anomaly \eqref{integratedWeyl}. 
To summarize the logic, we calculated  $W_{CFT}(g_{(0)})-W_{CFT}(\hat{g}_{(0)})=\lim\limits_{\bar{\rho}\rightarrow \infty}(S_{grav}[FG(\rho<\bar{\rho},x)]-S_{grav}[\widehat{FG}(\rho<\bar{\rho},x)])=\lim\limits_{\bar{\rho}\rightarrow \infty}(S_{grav}[\widehat{FG}(\rho^{\prime}<\mathcal{F}_{\bar{\rho}}(x^{\prime}),x^{\prime})]-S_{grav}[\widehat{FG}(\rho<\bar{\rho},x)])=S_L$.

In the top line in the bulk triangle, the actions and induced line elements are the same. 
It can be straightforwardly checked that the difference between the Brown-York stress tensor for the top left (of $FG(\rho,x)$ at fixed cut-off) 
and the Brown-York stress tensor for the bottom right (of $\widehat{FG}(\rho,x)$ at fixed cut-off) is given by the Liouville stress tensor of $S_L$ (with $\lambda=0$ as commented above). This was also calculated in \cite{Skenderis:2000in} and at the linearized level in \cite{deHaro:2000vlm}. 
It confirms holographically the anomalous stress tensor behavior, written as \eqref{LBmROtensor}, 
and as such provides a reformulation of the famous results of  \cite{Balasubramanian:1999re} in terms of the Liouville description.

	We first comment on previous discussions of the holographic Weyl anomaly and then proceed to discuss the implications of our calculation.

	\paragraph{Holographic Weyl anomaly}

	We compare to the original discussions of the non-integrated holographic Weyl anomaly.

	In the seminal work \cite{Henningson:1998gx}, the holographic Weyl anomaly was discussed in terms of the `logarithmic' divergence. It is logarithmic in the FG coordinate $r$ in \eqref{gind} and thus linear in our $\rho = -\log r$.
	The original argument considered a simultaneous change in $\hat g_{(0)}$ and 
	$\rho$ 
	\ali{
		\delta \hat g_{(0)} =  
		\phi \, \hat g_{(0)}, \qquad \delta \bar \rho = \frac{1}{2} 
		\phi   \label{Sktransf}
	} 
	with 
	$\phi$  
	infinitesimal and \emph{constant}. They 
	then proceeded to calculate $\delta S_{grav}$ in \eqref{hatFGmhatFG} with $\mathcal F_{\bar \rho}(x')$ given by $\bar \rho + \delta \bar \rho$, leading indeed to the action \eqref{SLholoWeyl} for constant and linear $\phi$, namely $\delta S_{grav} = \frac{c}{48 \pi} \int d^2 x \sqrt{-\hat g_{(0)}}   \phi \,  \hat R_{(0)}$,   
	produced by the difference in `logarithmic' divergent terms. 
	This represents the conformal Ward identity $\delta S_{grav}/\delta \phi$ 
	in  \eqref{Wardid}. Upon integration, it would lead only to \eqref{SLholoWeyl} without kinetic terms. To obtain the full Liouville action \eqref{SLholoWeyl} or integrated Weyl anomaly, it should be integrated for $\phi$ assumed non-constant. 
	The reason it is sufficient to work with constant $\phi$ for the derivation of the conformal Ward identity  
	is the following.  
	The structure of the Brown-Henneaux transformations \eqref{BrHexp} is such  
	that the simple 
	constant $\phi$ form of the transformation
	\ali{
		\rho' = \rho + \frac{1}{2} \phi, \qquad x'=x 
	} 
	is  
	equal to the leading behavior at small $\phi$ (when only retaining the powers that are necessary for obtaining \eqref{curlyFsimple} to first non-trivial order), or asymptotically.   
	Indeed all the coefficients $a^{(2j)}$ in the expansion consist of derivatives of $\phi$ and progressively higher powers in $\phi$. Therefore 
	the expansion in \eqref{BrHexp} refers not only to an expansion in $e^{-2\bar \rho}$ (near-boundary), but also in small $\phi$ (small fluctuations of the curly boundary \eqref{curlycutoff} around the fixed one), as well as a derivative expansion.  
	As a consequence, the exact, constant $\phi$ form of the curly boundary 
	\ali{
		\mathcal F_{\bar \rho}(x') = \bar \rho + \frac{1}{2} \phi   \label{curlyFsimple}
	}
	is also the correct form for either infinitesimal $e^{-2\bar \rho}$ \emph{or} infinitesimal $\phi(x)$. 
	This represents a significant simplification that can be exploited in the derivation of $\delta S_{grav}/\delta \phi$. 
	
	The transformation \eqref{Sktransf} 
	leads to the intuitive picture of holographic RG as moving from one constant $\bar \rho$ asymptotic cut-off to another constant $\bar \rho$ asymptotic cut-off within the same background geometry $\widehat{FG}$.   
	However, as we have discussed in our derivation of \eqref{SLholoWeyl}, it is  more correct to think of the latter cut-off as being non-constant or `curly',
	as pictured in Fig.~\ref{trianglecirclesfigure}.

	\paragraph{Horizon physics}

	We now return to  
	our finding that there are no contributions to the expected Liouville action from the upper bound of integration in $\rho$ in  Eq.~\eqref{Deltalbc}. 
	
	To recap, both our derivation of the holographic integrated Weyl anomaly \eqref{SLholoWeyl} and the original derivation of the holographic Weyl anomaly  
	involved the calculation of the difference in actions \eqref{hatFGmhatFG} along the vertical arrow in Fig.~\ref{trianglecirclesfigure}. 
	
	By the equality of Eq.~\eqref{FGmhatFGrefined} and Eq.~\eqref{hatFGmhatFG}, it follows immediately from the results \eqref{Liouholo} and \eqref{SLholoWeyl} that also the difference in actions \eqref{FGmhatFGrefined} along the diagonal arrow in Fig.~\ref{trianglecirclesfigure} has to equal the Liouville action 
	\ali{
		\lim_{\bar \rho \ra \infty} \left( S_{grav}[FG(\rho < \bar \rho,x^\mu)] - S_{grav}[\widehat{FG}(\rho < \bar \rho,x^\mu)] \right) = S_L . 
	}
	On the other hand, we know from Eq.~\eqref{Deltalbc} that the right hand side can only contain contributions from the lower bound of integration in $\rho$, which we have neglected to discuss so far. Indeed, the equality between \eqref{FGmhatFGrefined} and \eqref{hatFGmhatFG} rests on the equality of actions \eqref{equalityS} in the passive coordinate transformation step (horizontal arrow), which of course requires that the integration bounds are also accordingly changed. This was appropriately done for the upper bound of integration, but let us now also include the lower one to write \eqref{equalityS} more correctly as   
	\ali{
		S_{grav}[FG(\,\mathcal H_{\rho_+}(x) < \rho < \bar \rho,x^\mu)] = S_{grav}[\widehat{FG}(\rho_+ < \rho'< \mathcal F_{\bar \rho}(x'),x'^\mu)] . 
	}
	Again it is helpful to think of $\widehat{FG}$ as a Poincar\'e AdS geometry and $FG$ as a BTZ one. The lower bound of integration for the action evaluated on $\widehat{FG}$ would then be the Poincar\'e horizon, at $\rho' = \rho_+ \ra -\infty$. This maps under the Brown-Henneaux coordinate transformation to a horizon location in $FG$ that is dependent on the longitudinal coordinate  
	\ali{
		\rho = \rho\left(\rho_+,x'(\rho_+,x)\right) = \mathcal H_{\rho_+}(x) \label{curlyhorizon}
	}
	in the same way the boundary location was mapped in \eqref{curlycutoff} from a constant value in $FG$ to a non-constant value in $\widehat{FG}$. 
	This curly horizon depends on $x$ via the field $\phi(x)$.  

\begin{figure}[t]
		\centering
		\includegraphics[scale=0.4]{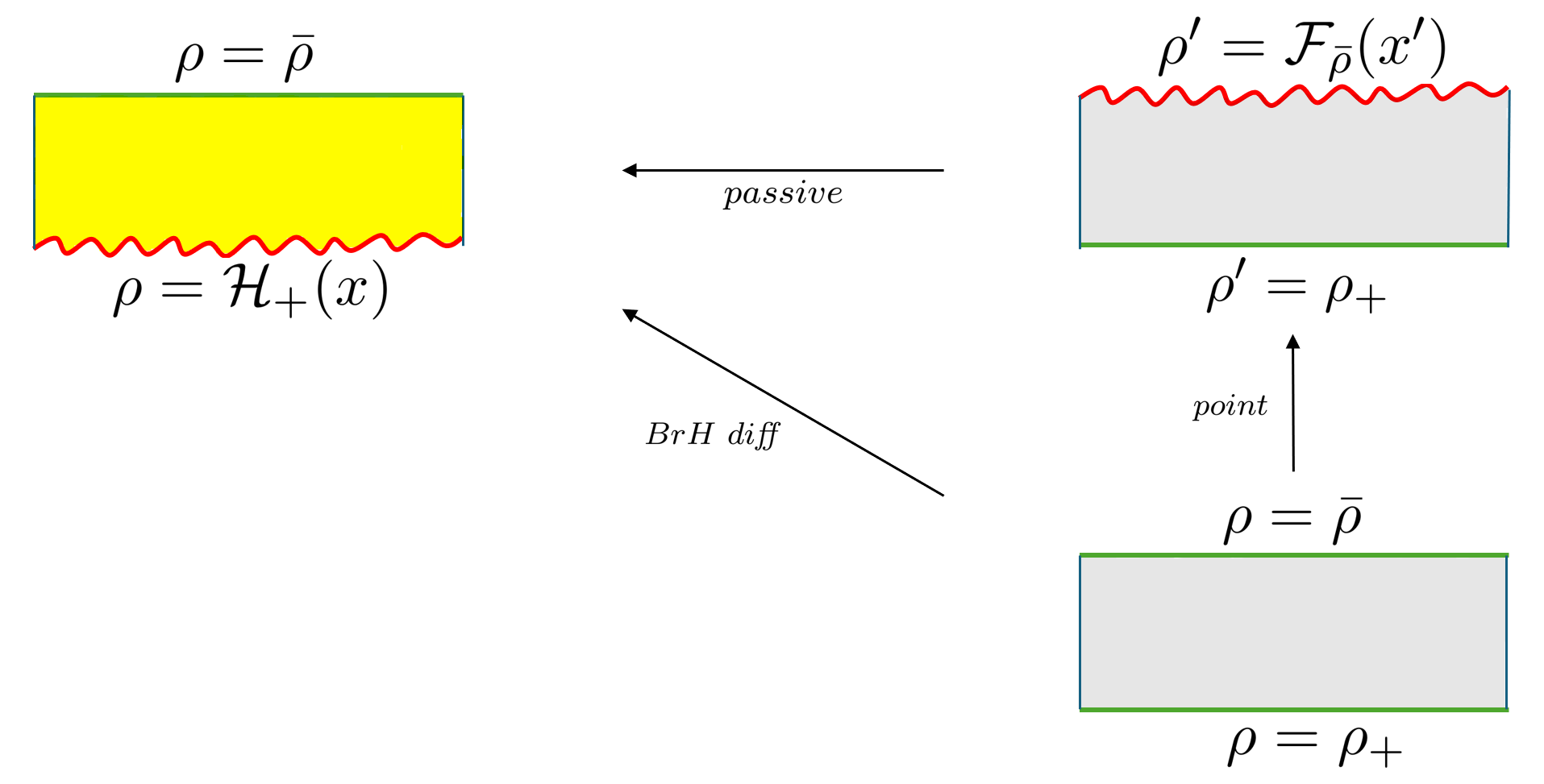}
		\caption{
Depiction of the bulk geometries that enter the calculation of the Liouville action $S_L$ from \eqref{WCFTsubtraction}, as in Fig.~\ref{trianglecirclesfigure}, but now also including the boundaries at the lower end of the radial bulk integration. Actions are the same in the top line, equal to $S + S_{L}$ compared to the starting point action $S$ (lower right corner). The red wiggly lines signify $\phi$-dependence, showing in the top line that the Liouville action encodes both asymptotic boundary physics (right) and horizon physics (left). It is helpful to think of the grey geometry $\hat G$ as the Poincar\'e solution and as the yellow geometry $G$ as BTZ, with $\rho_+$ the Poincar\'e horizon.			
	} 
		 \label{figbulkhorizon}
	\end{figure}

	This suggests that the Liouville theory \eqref{SLholoWeyl} has two interpretations. It describes the integrated Weyl anomaly physics of the asymptotic `curly boundary' in the geometry $\widehat{FG}$. But it also 
	must describe the horizon physics of the `curly horizon' in the geometry $FG$!  When we move to the braneworld discussion, this will mean that $S_L$ describes both the dynamics of an asymptotic brane in Poincar\'e AdS and the dynamics of a brane 
	approaching the Fefferman-Graham horizon in BTZ\footnote{We distinguish here between the horizon in FG coordinates and for example in Schwarzschild coordinates. Using the latter, the distinction between asymptotic boundary and horizon contributions is not as clear-cut as in FG coordinates.}. For exemplary calculations, we refer to Appendix \ref{appendix:B}. Both interpretations are illustrated in Fig.~\ref{figbulkhorizon}.

	\subsection{Asymptotic $3D$ braneworld} \label{subsholo}  
	
	In the previous subsection we have derived from a bulk perspective the integrated Weyl anomaly 
	\ali{
		Z_{CFT}(g_{(0)}) = Z_{CFT}(\hat g_{(0)}) e^{i S_L[\phi]}.   \label{integratedWeylrepeat}
	} 
	To discuss an asymptotic brane, we consider specifically the case where the tension is set to one ($T=1$), so that $S_{bwgrav} \equiv S_T-S_{ct}$ in \eqref{Sbw3D} vanishes. 
This is a natural choice for asymptotic braneworlds, because it ensures that the total action is free of power-law divergences: while $Z_{grav}$ is free of such divergences by construction due to the inclusion of counterterms, $S_{bwgrav}$ in the braneworld theory gives rise to divergences in the asymptotic limit unless the tension term 
	matches the 
		holographic counterterm. 
	We can then proceed to set free the CFT at the conformal boundary \cite{Compere:2008us} to obtain an asymptotic braneworld theory 
	\ali{
		Z_{bw} = \int \mathcal Dg_{(0)} \, Z_{CFT}(g_{(0)}) =  \int \mathcal Dg_{(0)} \, Z_{CFT}(\hat g_{(0)})  e^{i S_L[\phi]}.  \label{asbw}
	}
	The Liouville action \eqref{SLholoWeyl} takes the role of effective $2d$  gravitational action, with $c/(48\pi) = 1/(2\kappa_2)$   
	and thus 
	\ali{
		c &= \frac{12\pi}{\kappa}  
		\label{bulkbdypar3D}
	}
	providing the relation between the effective gravity ($c$) and bulk gravity parameters ($\kappa$). 
	
	In $2d$ quantum gravity, the integration over all $2d$ metrics $\mathcal D g_{(0)}$ represents an integration over gauge-inequivalent metrics. The physical degree of freedom in $g_{(0)}$ is the Weyl mode $\phi$, 
	and the overcounting due to $2d$ diffeomorphism invariance 
	is accounted for by a Faddeev-Popov determinant $\Delta_{FP}$ that can be written as a ghost theory $\int \mathcal D[b,c] \exp{i S_{gh}[b,c]}$ with central charge $c_{gh} = -26$ \cite{ZZ_Liouville_gravity,Erbin2015_2dGravityLiouville,Polyakov:1981rd}.
	In practice, we will be interested from a holographic perspective in the large $c$ $(\gg 26)$ limit, or from the braneworld perspective, in regimes where a saddle point approximation of the braneworld theory, with coupling constant $1/c$, is valid. Therefore, we will drop the determinant and effectively write 
	\ali{
		Z_{bw} = Z_{CFT}(\hat g_{(0)}) \int \mathcal D\phi \,  e^{i S_L[\phi]}.    \label{intZCFT} 
	}
	The integration is over the physical degree of freedom $\phi$ only, and $c_L=-c$ large justifies the suppression of the 
	determinant factor $\Delta_{FP}$. 
	
	Coming from a Dirichlet holography perspective, followed by setting the theory free, we have been working with finite $\phi$. 
	In traditional braneworld, it is standard to consider only small fluctuations. 
	When restricting over configurations close to a fixed brane, 
	the integration in \eqref{intZCFT} will be over small $\phi$ specifically, and the Liouville action is accordingly expanded to quadratic $\phi$. 
	The braneworld theory in \eqref{intZCFT} presents the asymptotic, $T=1$ case. Before generalizing to the non-asymptotic
	case, we make a few more remarks about 
	the appearance of Liouville theory in AdS$_3$ gravity. 
	
	Our holographic calculation of Liouville theory as \eqref{hatFGmhatFG} appeared first in \cite{Carlip:2005tz}. 
	While the calculation is identical,  
	we believe our interpretation to be different\footnote{We thank Rodolfo Panerai for pointing out to us that the Liouville action of \cite{Carlip:2005tz} should be understood in terms of the holographic Weyl anomaly of \cite{Henningson:1998gx}. See also \cite{Takayanagi:2018pml} and \cite{Suzuki:2022xwv}.}.   
	Namely, we interpret $\phi$ in \eqref{integratedWeylrepeat} as a free to choose boundary condition field. For $\phi=0$ or boundary condition $\hat g_{(0)}$ on the bulk metric, the bulk solution is of the form $\widehat{FG}$; for general $\phi$ or boundary condition $g_{(0)}=e^\phi \hat g_{(0)}$ 
	on the bulk metric, 
	it is $FG$. Making use of our bulk triangle representation of the BrH diffs  (see also \cite{Banados:1998gg}),  
	we  
	were able to map the difference \eqref{hatFGmhatFG} of two $\widehat{FG}$ actions with different boundaries  
	to the difference \eqref{FGmhatFGrefined} between the actions on bulk solutions that differ by a Weyl factor $e^{\phi}$ in their boundary condition. As such, we interpret the Liouville action resulting from \eqref{hatFGmhatFG} as the integrated Weyl anomaly of the CFT. 
	It is only in the `setting free' step \eqref{intZCFT} that $\phi$ becomes a dynamical field, in the sense that it is path integrated over. The setting free procedure is artificial, 
	making the field dynamical by hand to define a different theory, namely going from a Dirichlet theory to a Neumann theory.   
	The Weyl anomaly Liouville theory $S_L$ is therefore not to be confused (as also pointed out in \cite{Compere:2008us})  
	with other, effective Liouville theories in AdS$_3$ gravity \cite{Coussaert:1995zp}. 
	
	\section{$3D$ braneworld holography: $T\bar T$ set free} \label{sectTT}
	
	Based on our discussion in section \ref{sectWeyl} of asymptotic branes in $3D/2d$ holography, we consider in this section what happens when we bring the brane inwards into the bulk, by from now on allowing the location $\bar \rho$ of the brane to be a finite value. 
	
	The asymptotic holographic braneworld statement 
	\eqref{intZCFT} was obtained from setting free the integrated Weyl anomaly of the dual CFT \eqref{integratedWeyl}. We want to similarly construct the non-asymptotic holographic braneworld from setting free $T\bar T$ (subsection \ref{subsectTTfree}), with $T\bar T$ interpreted as a bulk extension of the 
	Weyl anomaly (subsection \ref{subsectTTWeyl} and \ref{subsectTTSL}).  
	To this end  
	we reconsider the set-up in Fig.~\ref{trianglecirclesfigure} with $\bar \rho$ non-asymptotic, and calculate the difference between the on-shell action for the top line 
	and the one for the bottom line. The braneworld discussion only starts in subsection \ref{subsectTTfree}. It is important to point out that in the preceding subsections, \ref{subsectTTWeyl} and \ref{subsectTTSL}, the Weyl mode is not yet `instructed' to be dynamic.

\subsection{Constructing the higher-derivative extension of the Liouville action $S_{\tilde L}$} \label{subsectTTWeyl} 
	
	\paragraph{The $\phi$ and $\tilde \phi$ fields }  
	Our strategy is to repeat the same steps as in the holographic derivation of the integrated Weyl anomaly $S_L$, essentially calculating the equivalent of \eqref{hatFGmhatFG}, but now for finite $\bar \rho$, 
	\ali{
		S_{grav}[\widehat{FG}(\rho'< \mathcal F_{\bar \rho}(x'),x'^\mu)] 
		- S_{grav}[\widehat{FG}(\rho < \bar \rho,x^\mu)] .  \label{hatFGmhatFGfinite}
	}     
	
	The function $\mathcal F_{\bar \rho}(x')$ again represents the 
	profile of the bulk direction $\rho'$ in the background $\widehat{FG}(\rho',x')$ that corresponds to the location $\bar \rho$ in the background $FG(\rho,x)$, in Fig. \ref{trianglecirclesfigure}. It is directly obtained from the Brown-Henneaux diff between the geometries $\widehat{FG}$ and $FG$, and by construction is such that the induced metrics 
	at the respective cut-offs are equal in the top line of the bulk triangle \eqref{bulktriangle}, 
	\ali{ 
		G_\mn(\bar \rho,x) dx^\mu dx^\nu = \p_\mu \mathcal F_{\bar \rho}(x') \p_\nu \mathcal F_{\bar \rho}(x') dx'^\mu dx'^\nu + \hat G_\mn( \mathcal F_{\bar \rho}(x'),x' ) dx'^\mu dx'^\nu. \label{equalityindmetrics}
	} 
	In Eq.~\eqref{curlycutoff}, the function $\mathcal F_{\bar \rho}(x')$ was given only to the necessary order to arrive at  
	equal induced metrics $e^{2\bar \rho}e^{\phi(x)} \hat g^{(0)}_\mn(x) dx^\mu dx^\nu$ to \emph{leading} asymptotic order. 
	To next order,
	it is given by 
	\ali{
		\mathcal F_{\bar \rho}(x') &= \bar \rho + \frac{1}{2} \phi(x') - \frac{1}{16} e^{-2\bar \rho} e^{-\phi(x')} \hat g_{(0)}^\mn(x') \p_\mu \phi(x') \p_\nu \phi(x') + \mathcal O(e^{-4\bar \rho})  \label{curlyF} \\
		&= \bar \rho + \tilde \phi(x').  \label{curlyFtildephi}
	} 
	This achieves equality of the induced metrics 
	in \eqref{equalityindmetrics} to order $\mathcal O(1)$, as can be checked using \eqref{FGp}, \eqref{FG2} and \eqref{g0phi}-\eqref{g2phi}.     
	The second line \eqref{curlyFtildephi} defines the mode $\tilde \phi$ to higher orders in $\phi$. 
	
	Plugging the  
	expression for $\mathcal F_{\bar \rho}(x')$ into \eqref{hatFGmhatFGfinite} will holographically calculate for us 
	\ali{
		W_{T\bar T}(g)  - W_{T\bar T}(\hat g) = S_{\tilde L}
		\label{Wttbarsubtraction}
	}
	with $g$ and $\hat g$ respectively the induced metric fields at $\rho = \bar \rho$ of $FG(\rho,x)$ and $\widehat{FG}(\rho,x)$. 
	The result will be an action for $\tilde \phi$ or 
	for the field $\phi$ labeling the Brown-Henneaux diffs, 
	and will reduce to the Liouville action $S_L$ for the (asymptotic) conformal mode $\phi$ in the $\bar \rho \ra \infty$ limit. It is this action $S_{\tilde L}$ 
	that will acquire 
	the interpretation of braneworld gravity upon setting free $g$, 
	and that we want to derive explicitly from this bulk argument.  
	 
	The above 
	outlines a systematic procedure to obtain the $S_{\tilde L}$ action   
	in a perturbative expansion away from the asymptotic boundary, making use of the Brown-Henneaux diffs in the series expansion form \eqref{BrHexp}. 
	It can be done in full generality, but becomes quite tedious quickly. We therefore restrict $\widehat{FG}$ to be a Poincar\'e solution $Poinc$, 
	\ali{
		\hat g_{(0)} = \eta, \qquad \hat g_{(2)} = \hat g_{(4)} = 0 .  \label{Poincreference}
	}
	Indeed, this is a natural choice for later comparison to braneworld literature, where the bulk is typically taken to be vacuum AdS. 

	Similar to Eq.~\eqref{SgravFGgeneral}, we can compute the action $S_{grav}[Poinc(\rho < F(x),x)]$ evaluated on the Poincar\'e solution with $\rho$ integrated up to a general boundary $F(x)$. The result is given in Appendix \ref{appendix:D} (the Appendix also contains the more general $S_{grav}[Ban(\rho < F(x),x)]$).   
From it, we can read off that 
the result for \eqref{hatFGmhatFGfinite} becomes 
\begin{align}
	S_{\tilde{L}}=\frac{1}{2\kappa}\int d^2x  \sqrt{-\hat{g}_{(0)}}\Bigg\{&\tilde{\phi}\hat{R}_{(0)}+2e^{2(\bar\rho + \tilde \phi)}+2(\partial \tilde{\phi})^2 
	-2e^{2(\bar\rho + \tilde \phi)}\sqrt{1+ e^{-2(\bar\rho + \tilde \phi)}(\partial \tilde{\phi})^2}
	\nonumber
	\\
	&+\frac{\partial^{\mu} \tilde{\phi}\left[-2e^{-2(\bar\rho + \tilde \phi)}\partial_{\mu}\tilde{\phi}(\partial \tilde{\phi})^2+e^{-2(\bar\rho + \tilde \phi)}\partial_{\mu}(\partial \tilde{\phi})^2\right]}{1+e^{-2(\bar\rho + \tilde \phi)}(\partial \tilde{\phi})^2}\Bigg\}
	\label{StildeLtildephiresult}
\end{align} 
with $(\p \tilde \phi)^2 \equiv  \hat{g}_{(0)}^\mn \p_\mu \tilde \phi \p_\nu \tilde \phi$. 
It depends on the fixed brane location $\bar \rho$ both explicitly and implicitly through $\tilde \phi$, and on the asymptotic Weyl mode $\phi$ through $\tilde \phi$. It can be written out perturbatively in terms of $\phi$. 
	To first corrected order in $\epsilon \equiv e^{-\bar \rho}$, the action $S_{\tilde L}$ is then given by 
	\ali{
		S_{\tilde{L}}&=\frac{c}{48\pi}\int d^2x\sqrt{-\hat{g}_{(0)}}\left\{ \phi \hat{R}_{(0)}+\frac{1}{2}(\p \phi)^2 \right. \nonumber \\ 
		&\qquad \left. -\frac{\epsilon^2}{8e^{\phi}}\Bigg[\frac{3}{4}(\p \phi)^4-\hat g_{(0)}^\mn \partial_{\nu} \phi\partial_{\mu}(\p \phi)^2+ \hat R_{(0)}(\p \phi)^2  
		\Bigg]+\mathcal{O}(\epsilon^4)\right\}  \label{StildeLPT}
	}
	with $(\p \phi)^2=\hat{g}_{(0)}^{\mu\nu}\partial_{\mu}\phi\partial_{\nu}\phi$, and a total derivative was dropped. The notation in terms of $\hat g_{(0)}$ was kept to see the asymptotic Liouville form appear in the first line, but can be replaced by $\eta$ notation to 
	\ali{
		S_{\tilde{L}}=&\frac{c}{48\pi}\int d^2x\Bigg\{\frac{1}{2}(\p \phi)^2 +\frac{\epsilon^2}{32}e^{-\phi}(\p \phi)^4-\frac{\epsilon^2}{8}e^{-\phi}(\p\phi)^2 \eta^\mn \partial_{\mu}\partial_{\nu}\phi+\mathcal{O}(\epsilon^4)\Bigg\}. 
		\label{StildeLphiresult} 
	}

Equations \eqref{StildeLtildephiresult} and \eqref{StildeLphiresult} are our main results for the action $S_{\tilde L}$ defined in \eqref{Wttbarsubtraction}. 
	The $\phi$ field has the physical interpretation of being the asymptotic Weyl mode. It provides the natural description of $S_{\tilde L}$ from the AdS/CFT perspective, 
	making use of BrH diffs.
	For braneworld interpretations, it is the $\tilde \phi$ mode that is the natural object to consider, as it will become the `radion' in the language of e.g.~\cite{Geng:2022slq,Geng:2022tfc}.

	\paragraph{Making use of finite BrH diffs} 
	
	Constructing the function $\mathcal F_{\bar \rho}$ order by order in $e^{-2\bar \rho}$ by building the series representation of the BrH diffs in \eqref{BrHexp} becomes quite involved. 
	A closed-form expression for \eqref{BrHexp}  
	is known for the transformation between a Poincar\'e AdS solution and a general Banados solution \cite{Roberts:2012aq}. 
	So if we 
consider from now on $\widehat{FG}$ to be a Poincar\'e solution $Poinc$ as well as $FG$ to be a Banados solution $Ban$, 
	we can aim to obtain a closed-form expression for the curly brane location $\mathcal F_{\bar \rho}(x')$. 
	Our starting point in \eqref{FG1} becomes the Poincar\'e metric 
	\ali{
		\hat G(X)dX^2 = d\rho^2 + e^{2\rho} (-2 df d \bar f).  
	} 
	It is transformed into the Banados metric 
	\ali{
		&G(X) dX^2 = d\rho^2 + e^{2\rho} e^\phi (-2 df d\bar f) - \frac{6}{c} t_{ff}^L df^2  - \frac{6}{c} t_{\bar f \bar f}^L d\bar f^2 -\frac{18}{c^2}\frac{e^{-2\rho}}{e^{\phi}}t_{ff}^L t_{\bar f \bar f}^L \,  df d\bar{f} \label{Ban} 
	}
	by the finite Brown-Henneaux diff 
	\ali{
		\begin{split}
			\rho'(\rho,x) &= \rho + \frac{1}{2}\phi(x) + \log\left( 1 + \frac{1}{16} e^{-2\rho} e^{-\phi(x)} \eta^\mn \p_\mu \phi \p_\nu \phi \right)  \\ 
			x'^\mu(\rho,x) &= x^\mu + \frac{e^{-2\rho} e^{-\phi(x)} \eta^\mn \p_\nu \phi}{4+\frac{1}{4}e^{-2\rho} e^{-\phi(x)} \eta^\mn \p_\mu \phi \p_\nu \phi}  
		\end{split}  \label{BrHRoberts}	
	}
	with the first line also taking the succinct form $e^{2\rho'} = e^{2\rho} e^{\phi} + \eta^\mn \p_\mu \phi \p_\nu \phi /16$.  
	This is the 
	BrH diff of \cite{Roberts:2012aq}, given there in terms of $f$ and $\bar f$, and rewritten here in terms of $\phi$, using  \eqref{phisol}, i.e.
	\ali{
		\phi = -\log \left( \frac{\p f}{\p z} \frac{\p \bar f}{\p \bar z} \right)  
		\label{philog}
	}
	and \eqref{tL} for the Liouville stress tensor $t_\mn^L[\phi]$. With this Weyl factor we have restricted to flat slicing of the bulk, and the Banados geometry \eqref{Ban} takes the more familiar form 
	\ali{
		d\rho^2 + e^{2\rho} (-2 dz d\bar z) - \frac{1}{2} \{f,z\} dz^2  - \frac{1}{2} \{\bar f,\bar z\} d\bar z^2 - \frac{1}{8} e^{-2\rho} \{f,z\}\{\bar f,\bar z\} dz d\bar z. 
	} 
	The bulk triangle representation of the diffeomorphism is\footnote{Let us remark that to compare to the notation of \cite{Roberts:2012aq}, our $\rho$ and $x^\mu = (f,\bar f)$ are their $(-\log Z)$ and $(f_+,f_-)$ respectively, and our $\rho'$ and $x'^\mu = (y,\bar y)$ are their $(-\log u)$ and $(y_+,y_-)$.}  
	\begin{equation}
		\begin{tikzcd}[row sep=3cm, column sep=4cm]
			\begin{array}{c}  
				X = (\rho, f, \bar f) \\[8pt] 
				G(X)\, dX^2 \;=\; Ban(\rho, x^\mu)   
			\end{array}
			&
			\begin{array}{c}  
				X' = (\rho', y, \bar y) \\[8pt]  
				\hat G(X')\, dX'^2 \;=\; Poinc(\rho', x'^\mu) 	
			\end{array}
			\arrow[l, "\;\; passive \;\;"]
			\\
			&
			\begin{array}{c}  
				X = (\rho, f, \bar f) \\[8pt] 
				\hat G(X)\, dX^2 \;=\; Poinc(\rho, x^\mu)  
			\end{array}
			\arrow[u, "\;\; point \;\;"] 
			\arrow[ul, "\;\; BrH \,\, di\mathit{ff} \;\;"]
		\end{tikzcd}
		\label{bulktriangle3}
	\end{equation}
	
	The transformation of $\rho$ in \eqref{BrHRoberts} is of the form $\rho' = \rho + \psi(\rho,x)$ with  
	\ali{
		\psi = \frac{1}{2}\phi + \log\left( 1 + \frac{1}{16} e^{-2\rho} e^{-\phi} \eta^\mn \p_\mu \phi \p_\nu \phi \right).  \label{psi}
	}
	The curly boundary location in the Poincar\'e background  
	consequently takes the form \eqref{curlyFtildephi} 
	with 
	\ali{
		\tilde \phi \equiv \psi\left( x(\bar \rho,x') \right).  
	}
	Because we only are able to determine the inverse BrH diffs 
	perturbatively, we in fact only know $\tilde \phi$ and thus $\mathcal F_{\bar \rho}(x')$ perturbatively\footnote{
		To first order, to repeat from \eqref{curlyF}, \ali{
			\tilde \phi = \frac{\phi}{2} - \frac{1}{16} e^{-2\bar \rho} e^{-\phi} \eta^\mn \p_\mu \phi \p_\nu \phi + \mathcal O(e^{-4\bar \rho}). \label{phitildeofphi} 	
		}
	}, for either large $\bar \rho$ or small $\phi$ or $\tilde \phi$.  
	But the advantage of working with \eqref{BrHRoberts} is that each term in the expansion of the BrH diff can be read off directly, allowing to expand the action \eqref{StildeLtildephiresult} systematically. 

Next, we discuss yet another mode $\sigma$, 
defined as the Weyl mode of the induced metric $g$,  
which will be the most natural one from a finite holography or $T\bar T$ perspective.

\subsection{$T\bar T$ in terms of its Weyl mode: a $T\bar T$-like flow for the action $S_{\tilde L}$} \label{subsectTTSL}	
	
	\paragraph{The $\sigma$ field } 
	So far we have naturally extended the holographic Weyl anomaly calculation further into the bulk in order to obtain $S_{\tilde L}$ in \eqref{Wttbarsubtraction} either as a function of $\tilde \phi$, the fluctuation of the boundary's radius \eqref{curlyFtildephi}, or $\phi$, the 
		Weyl mode of the conformal boundary metric $g_{(0)}$.  
	The reason for this strategy is that it allows us to exploit the  
	knowledge of 
	AdS$_3$ geometry and its BrH diffs.  
	However, neither of these fields is the Weyl mode that we want to set free 
	to obtain a braneworld theory \eqref{bwholo}. To set the induced boundary metric $g$ free in $W_{T\bar T}(g)$, we want to write it as 
	\ali{
		g_\mn = e^\sigma \hat g_\mn   \label{sigmadef}
	}
	and set free the 
		degree of freedom $\sigma$ of the induced metric $g$  in $S_{\tilde L}[\sigma; \hat g]$, in a notation analogous to the asymptotic \eqref{2Dmetric} and \eqref{integratedWeyl}. 	
	Locally, there is a two-dimensional passive coordinate transformation that brings the metric in this form \eqref{sigmadef}. 
In section \ref{subsholo}, the asymptotic \eqref{g0phi} and \eqref{integratedWeylrepeat} 
were taken as starting point to obtain a set free CFT as braneworld theory in \eqref{intZCFT}, with the Weyl mode $\phi$ of the conformal boundary metric path integrated over. Following a completely analogous setting free argument, but starting from respectively \eqref{sigmadef} and \eqref{Wttbarsubtraction} written in the form $Z_{T\bar T}(g) = Z_{T\bar T}(\hat g) e^{i S_{\tilde L}[\sigma]}$, we will integrate over the Weyl mode $\sigma$ of the induced metric $g$ to set $T\bar T$ free.

	To determine 
	the action $S_{\tilde L}$ as a function of $\sigma$, let us  consider the Hamiltonian constraint of AdS$_3$ gravity, written as a Hamilton-Jacobi equation for $S$ an  
	on-shell gravitational action $S_{grav}$,   
	\ali{
		4\kappa \frac{1}{\sqrt{-g}} g_{\mn} \frac{\delta S}{\delta g_{\mn}}=R - \frac{(2\kappa)^2}{g} \left(-\frac{\delta S}{\delta g_{\mn}} \frac{\delta S}{\delta g^{\mn}} - \left(g_{\mn}\frac{\delta S}{\delta g_{\mn}}\right)^2 \right)  
		. \label{HJeq}
	} 
	By writing it in terms of the Brown-York stress tensor $T^{BY}_\mn = -\frac{4\pi}{\sqrt{-g}} \frac{\delta S}{\delta g^\mn}$, 
	\ali{
		&\frac{1}{2\pi} T^{\mn}_{BY}g_{\mn}=\frac{1}{2 \kappa } R + \frac{\kappa}{8\pi^2} \left( T^{\mn}_{BY} T^{\ab}_{BY} g_{\alpha\mu} g_{\beta\nu} - (T^{\mn}_{BY}g_{\mn})^2 \right)   , 
		\label{HJeq2pi}
	}
	it directly matches the $T\bar T$ trace flow equation 
	\ali{
		&\vev{T^\mu_\mu}=-\frac{c}{12} R +  t \,\,  \vev{\mathcal O_{T\bar T}}_{g} 
		, \qquad  \mathcal O_{T\bar T} \equiv  T^{\mn} T^{\ab} g_{\alpha\mu} g_{\beta\nu} - (T^{\mn}g_{\mn})^2 
		\label{vevtraceflow}
	}
	for the $W_{T\bar T}(g)$ theory, i.e.~$T_\mn^{BY}$ has a dual interpretation as 
	(minus) the $T\bar T$ stress tensor $\vev{T_\mn} = \frac{4\pi}{\sqrt{-g}} \frac{\delta W_{T\bar{T}}}{\delta g^\mn}$. 
	This is the result of \cite{McGough:2016lol}, see also \cite{Marolf18}, for the flow \eqref{TTflow} with 
\ali{
	t = - \frac{\kappa l}{4\pi}. 
}
	Upon writing the metric as \eqref{sigmadef}, the curvature term will take the form $\sqrt{-g} \, R = \sqrt{-\hat g} \hat R - \sqrt{-\hat g} \hat g^\mn \p_\mu \p_\nu \sigma$   
	and the stress tensors will change as follows.  In complete analogy with the 
	derivation of \eqref{LBmROtensor} or \eqref{deltalogZCFT}, using the definition of $\sigma$ in \eqref{sigmadef} and the definition of $S_{\tilde L}$ in \eqref{Wttbarsubtraction}, the Brown-York stress tensor associated with the upper left ($FG$) corner of Fig.~\ref{trianglecirclesfigure} will differ from the one associated with the lower right ($\widehat{FG}$) corner by a $\sigma$-dependent contribution 
	\ali{
		\vev{T_\mn}_{e^\sigma \hat g} = \vev{\hat T_\mn}_{\hat g} + t_\mn^{\tilde L}[\sigma]    \label{TmTL}
	} 
	with $t^{\tilde L}_\mn$ defined as $\frac{4\pi}{\sqrt{\hat g}} \frac{\delta S_{\tilde L}}{\delta \hat g^\mn}$ and $\vev{\hat T_\mn}_{\hat g}$ as $\frac{4\pi}{\sqrt{\hat g}} \frac{\delta W_{T\bar T}(\hat g)}{\delta \hat g^\mn}$. 
	The Hamilton-Jacobi equation \eqref{HJeq} becomes 
	\ali{
		&(\vev{\hat T^\mn}+t^\mn_{\tilde L}) \hat g_{\mn}= \nonumber
		\\
		&\frac{c}{12} (\hat \Box \sigma-\hat R) + t \, e^{-\sigma} \left( (\vev{\hat T^\mn}+t^\mn_{\tilde L})(\vev{\hat T^\ab}+t^\ab_{\tilde L})\hat g_{\alpha\mu} \hat g_{\beta\nu} - \left((\vev{\hat T^\mn}+t^\mn_{\tilde L}\right) \hat g_\mn)^2 \right).  \label{traceflow}
	} 
	The new flow takes the same form as the original flow equation \eqref{vevtraceflow}, 
	\ali{
		&(\vev{\hat T^\mn}+t^\mn_{\tilde L}) \hat g_{\mn}=\frac{c}{12} (\hat \Box \sigma - \hat R) +  t \, e^{-\sigma} \,  \vev{\mathcal O_{T\bar T}}_{\hat g}  
		\label{ourTTflow} 
	}
	but with a coupling term that is modified by the factor $e^{-\sigma}$ compared to the standard $T\bar T$ one, $t \, \mathcal O_{T\bar T} \ra t e^{-\sigma} \mathcal O_{T\bar T}$. As such it describes a $T\bar T$-\emph{like} deformation, with the emphasis on `like' because the deformation depends on $\sigma$ in this non-trivial way. We note that such a rescaled deformation also appears in the context of the conformal boundary condition problem in \cite{Allameh:2025gsa}. 
	Indeed, a close relation between the NBC and CBC problem is to be expected in $3D/2d$, the first defined as $\int Dg Z_{T\bar T}$ (with $\sigma$ the physical mode)  and the second by a Laplace transform of $Z_{T\bar T}$. 
	Integrating the flow \eqref{ourTTflow} should give rise to a $\sigma$-dependent $W_{T\bar T}$, that substituted into \eqref{Wttbarsubtraction} gives an expression for $S_{\tilde L}$ in terms of $\sigma$, which is then to be set free by integrating over $\sigma$.
	
We restrict to a flat reference metric $\hat g$ (that is, $\hat R = 0$). 	
	To proceed, we decouple the flow by assuming a vacuum state $\vev{\hat T_\mn}_{\hat g} = 0$,  
	equivalent to the vanishing of the Brown-York stress tensor on the Poincar\'e background.  
	Then the trace flow equation \eqref{traceflow} only governs the `Liouville' $\sigma$ part,  
	\ali{
		&t_{\tilde L}^{\mn}\hat g_{\mn}=\frac{c}{12} \hat \Box \sigma + t e^{-\sigma} \left( t_{\tilde L}^{\mn} t_{\tilde L}^{\ab} \hat g_{\alpha\mu} \hat g_{\beta\nu} - (t_{\tilde L}^{\mn} \hat g_{\mn})^2 \right)   
		 .  \label{traceflowLiou} 
	} 
	This shows that $S_{\tilde L}$ is a $T\bar T$-like deformed Liouville theory 
	\ali{
		S_{\tilde L} \equiv S_{\tilde L}^{(t)} 
	}
	whose stress tensor $t^{\tilde L}_\mn = \frac{4\pi}{\sqrt{\hat g}} \frac{\delta S_{\tilde L}}{\delta \hat g^\mn}$ satisfies the $T\bar T$-like trace flow equation 
	\ali{
		t_{\tilde L \, \mu}^\mu=\frac{c}{12}\hat \Box \sigma + t \, e^{-\sigma} \mathcal O^{\tilde L}_{T\bar T,\hat g}, \qquad \mathcal O^{\tilde L}_{T\bar T,\hat g} \equiv t_{\tilde L}^\mn t_{\tilde L}^\ab \hat g_{\alpha\mu} \hat g_{\beta\nu} - (t_{\tilde L}^\mn \hat g_\mn)^2 . \label{traceflowliou} 
	}
	In the absence of the deformation or `asymptotically', $t=0$, this theory reduces to the timelike Liouville theory describing the holographic Weyl anomaly, 
	\ali{
		S_{\tilde L}^{(0)} = S_L. \label{seedtheory}  
	}
	The trace flow equation is the form of the $T\bar T$ flow that naturally arises in holography from the gravitational Hamilton-Jacobi equation. There is a standard argument for deducing from it the actual flow equation for the $T\bar T$ action, which can be found e.g.~in \cite{Marolf18}.  
	It assumes that the deformed theory has only one mass scale $\mu = 1/\sqrt{t}$, 
	and that as such the scaling variation of the action can be written as $\mu \frac{dS}{d\mu} = -2 t \frac{dS}{dt} = -\frac{1}{2\pi} \int d^2 x \sqrt{-g} \,  T^\mu_\mu$ for $\delta S_{QFT} = -\frac{1}{4\pi} \int T^\mn \delta g_\mn \sqrt{-g}$,  
	such that a trace flow equation of the form $T^\mu_\mu = 4 \pi \, t \, \mathcal O_{T\bar T}^{(t)}$ implies the flow equation for the action to be $\frac{dS}{dt} = \int d^2 x \sqrt{-g} \, \mathcal O_{T\bar T}^{(t)}$. 
	Applied to our trace flow, this argument implies the flow equation 
	\ali{
		\frac{d}{dt} S_{\tilde L}^{(t)} &= \frac{1}{4\pi} \int d^2 x \sqrt{-\hat g} \,  e^{-\sigma} \mathcal O^{\tilde L}_{T\bar T,\hat g}, \qquad S_{\tilde L}^{(0)} = S_L \label{actionflow} 
	}
	with $t = -\kappa l/(4 \pi)$. 
	We argue that the anomaly term $\hat \Box \sigma$ does not contribute to the right hand side of $\frac{dS_{\tilde L}}{dt}$ because it is a total derivative, and its effect is instead captured in the seed theory \eqref{seedtheory}.  
	More precisely, let us first use the seed theory with $\hat R = 0$, which is then just the Lorentzian timelike free boson 
	\ali{
		S_{\tilde F}^{(0)} &= \frac{c}{48\pi} \int d^2 x \sqrt{-\hat g} \frac{1}{2} \hat g^\mn \p_\mu \sigma \p_\nu \sigma  \, .   
	}  
	The corresponding flow $\frac{d}{dt} S_{\tilde{F}}^{(t)} = \frac{1}{4\pi} \int d^2 x \sqrt{-\hat g} \,  e^{-\sigma} \mathcal O^{\tilde{F}}_{T\bar T,\hat g}$ for the action $S_{\tilde{F}}\equiv S_{\tilde{F}}^{(t)} $ can be straightforwardly solved using the methods of \cite{Cavaglia:2016oda,Bonelli:2018kik,Leoni:2020rof} 
to obtain 
	\ali{
		S_{\tilde F} &= \frac{1}{4\pi} \int d^2 x \sqrt{-\hat g} \,\, e^{\sigma} \, \frac{1 - \sqrt{1 -  \, \frac{t \, c}{12}  e^{-\sigma} \hat g^\mn \p_\mu \sigma \p_\nu \sigma}}{t} . \label{StildeLsigma}
	} 
	The obtained Lagrangian satisfies the trace flow equation \eqref{traceflowliou} for the deformed Liouville stress tensor without the $\hat \Box \sigma$ contribution, 
	\ali{
		t_{\tilde F \, \mu}^\mu = t \, e^{-\sigma} \mathcal O^{\tilde F}_{T\bar T,\hat g} \, . \label{here}  
	}
	
	There is no $\hat \Box \sigma$ contribution in \eqref{here} because we started from a seed theory without the $\sigma \hat R$ term. The corresponding trace of the stress tensor vanishes, as can be seen from the  first two terms in the Liouville stress tensor expression \eqref{tL}. It is the final, third term in that expression that comes directly from 
	variation with respect to $\hat g_\mn$ in 
	the $\sigma \hat R$ term in the Liouville action, 
	and that will produce a $\hat \Box \sigma$ contribution in the Liouville trace flow. 
When including the 	$\sigma \hat R$ term, it is much harder to obtain a closed expression   
	\cite{Bonelli:2018kik}, but to first order starting from the Liouville seed theory 
	\ali{
		S_{\tilde L}^{(0)} &= \frac{c}{48\pi} \int d^2 x \sqrt{-\hat g} \left(\sigma \hat R + \frac{1}{2} \hat g^\mn \p_\mu \sigma \p_\nu \sigma \right)  
	}   
	gives rise to the deformed theory 
\begin{align}
	S_{\tilde{L}}=S_L+\frac{t}{16\pi}\left(\frac{c}{12}\right)^2 \int d^2x \sqrt{-\hat{g}}\, e^{-\sigma}\Bigg(&\frac{1}{2}(\partial \sigma)^4-2\hat{\Box}\sigma(\partial \sigma)^2\Bigg)+\mathcal{O}(t^2). \label{StildeLsigmat1} 
\end{align}	
Here indices are raised and lowered with $\hat g_\mn$, i.e. $(\partial \sigma)^2=\hat{g}^{\mu\nu}\partial_{\mu}\sigma \partial_{\nu}\sigma$. 
An additional $\hat{\Box}\sigma(\partial \sigma)^2$ term is present in the deformed Liouville theory $S_{\tilde L}$ compared to the deformed free boson theory $S_{\tilde F}$. 
	The new contributions 
restore	the missing $\hat \Box \sigma$ in the trace flow \eqref{traceflowliou} at zeroth order in $t$.  
The action \eqref{StildeLsigmat1} satisfies the action flow \eqref{actionflow} (by construction of the first order Lagrangian as $L_{\tilde L}^{(1)} \sim e^{-\sigma} \mathcal O_{T\bar T,\hat g}^{L}$ with the Liouville stress tensor components given in \eqref{tL})  
as well as the trace flow \eqref{traceflowliou}, both to first order in $t$. Higher orders can be added by solving the flow order by order.

We refer to Appendix \ref{AppTTbar} for more details on the derivation of the above deformed actions.

\paragraph{Summary on different modes}

Let us summarize the results of subsections \ref{subsectTTWeyl} and \ref{subsectTTSL} deriving $S_{\tilde L}$. Different strategies lead to different incarnations of $S_{\tilde L}$, meaning written in terms of different relevant Weyl fields $\phi$, $\tilde \phi$ or $\sigma$. In terms of $\sigma$, $S_{\tilde L}$ should be the $T\bar T$-like deformed Liouville theory, from a Hamilton-Jacobi argument. It can be constructed order by order by solving the Hamilton-Jacobi equation or $T\bar T$-like flow equation. 
A closed expression can be obtained for the related problem in which the seed theory is taken to be a (timelike) free boson such that $\hat \Box \sigma = 0$ in the trace flow.   
The BrH diffeomorphism-based 
strategy extending the Weyl anomaly calculation directly further into the bulk 
gives rise to an action in terms of the asymptotic Weyl mode $\phi$, in an asymptotic expansion. 
It is also naturally expressed in terms of the field $\tilde \phi$, 
which marks the location of the wiggly boundary in Poincar\'e AdS to which the constant $\bar \rho$ boundary corresponds. 
In the case of $\phi$ being of the particular form $\phi = -\log(f' \bar f')$, the relation between $\tilde \phi$ and $\phi$ can be computed systematically thanks to the closed form expression in \eqref{psi}.

In some sense, the diff-based strategies can be seen as the AdS$_3$ bulk geometrically solving the $T\bar T$-like flow equation for us. To see this explicitly we would require the order by order relation between $\sigma$ and $\phi$ (which we have not constructed), but let us discuss two ways in which this becomes apparent. 

First of all, it is clear that the structure of the obtained 
expressions for $S_{\tilde L}$ in terms of $\phi$ and $\tilde \phi$ is of the correct form to compare to the $T\bar T$-like deformed Liouville theory \eqref{StildeLsigmat1}. Namely, the expansion parameter is the same, as we will now discuss. 
The expansion in the result \eqref{StildeLphiresult} 
is in $\epsilon^2 \equiv e^{-2\bar \rho}$ and the background metric for the theory was fixed to $\hat g^{(0)}_\mn = \eta_\mn$. The expansion parameter in \eqref{StildeLsigmat1} is $(t \, c)$ or the $T\bar T$ coupling times the central charge, which in bulk language is a number independent of $\kappa$. The background metric in the latter case is the induced metric at $\rho = \bar \rho$ in the Poincar\'e bulk, $\hat g_\mn = e^{2\bar \rho} \eta_\mn$. As usual in $T\bar T$ holography, it is a matter of choice to include the factor $e^{2\bar \rho}$ in the background metric or in the identification of the $T\bar T$ coupling. 
To compare to the action \eqref{StildeLphiresult} explicitly, one would make the latter choice (i.e.~$\hat g_\mn = \eta_\mn$) and the expansion 
would indeed be in   
\ali{
	t \, c  = -3 \, e^{-2\bar \rho} \label{tcbarrho} 
} 
which is perhaps the more familiar relation for the $T\bar T$ coupling from \cite{McGough:2016lol}. 

Secondly, we make the following observation in comparing  
the different forms that we obtained for the action $S_{\tilde L}$. To linear order in the near-boundary expansion, the $S_{\tilde L}$ action in terms of $\phi$ given in \eqref{StildeLphiresult} takes the same form as the $T\bar T$-like deformed Liouville in terms of $\sigma$.

\subsection{Setting $T\bar T$ free}  \label{subsectTTfree}

In the previous subsections we have discussed the theory with action $S_{\tilde L}$, 
which we defined as 
\ali{
	Z_{T\bar T}(g) = Z_{T\bar T}(\hat g) e^{i S_{\tilde L}[\sigma]}.   \label{ZTTbar}
}
To obtain the braneworld theory of the finite 
brane according to \eqref{bwholo}, we should include the  
extra terms $S_T - S_{ct}$ and then set the $T\bar T$ theory free.  
As discussed in the introduction of section \ref{sectWeyl}, the extra terms produce area contributions $\sqrt{-g} \lambda$, which from the $\sigma$-theory perspective are Liouville potential contributions $\sqrt{-\hat g} \lambda e^\sigma$, with the 2d cosmological constant determined by the tension of the brane as $\lambda = 2(1-T)$.  
We restrict for simplicity to the theory with the tension parameter set to the value $T=1$. This is the natural choice for later comparison to traditional braneworld, in which $\hat g$ takes on the interpretation of a saddle 
satisfying the NBC (setting $T=1$ for $\hat g$ flat).  
With this choice there are no extra area term contributions and the theory that is set free is $Z_{T\bar T}(g)$. 
It reduces asymptotically to the CFT set free discussed in section \ref{subsholo}. 
For $T=1$, the braneworld theory from setting $T\bar T$ free is given by 
\ali{
	Z_{bw} &= 
	\int \mathcal Dg \, Z_{T\bar T}(g) = \int \mathcal Dg \, Z_{T\bar T}(\hat g) e^{i S_{\tilde L}[\sigma]}   
}
with 
\ali{
	c &= \frac{12\pi}{\kappa} , \qquad t =- \frac{\kappa}{4\pi}  
}
providing the relation between the effective gravity ($c, t$) and bulk gravity parameters ($\kappa, \bar \rho$) (and $l=1$). Compared to the asymptotic braneworld in \eqref{asbw}, there is an extra parameter $t$ on the boundary side of the duality for effectively measuring the location $\bar \rho$ of the brane on the bulk side. 

As in \eqref{intZCFT}, 
we rewrite the path integral over $g$ in terms of  
a path integral over the only physical degree of freedom being its Weyl mode $\sigma$, defined in \eqref{sigmadef}, and moreover suppress the appearance of the ghost action for large $c$ considerations\footnote{The argument followed is identical to that in section \ref{subsholo} deriving \eqref{intZCFT} starting from \eqref{g0phi} and \eqref{integratedWeylrepeat}, 
	now for \eqref{sigmadef} and \eqref{ZTTbar}.},
\ali{
	Z_{bw} &= Z_{T\bar T}(\hat g)  \int \mathcal D \sigma \,  e^{i S_{\tilde L}[\sigma]}.  \label{Zbw3d} 
}   
This is our conjectured $3D/2d$ braneworld theory for non-asymptotic branes or general brane location $\bar \rho$,  with $S_{\tilde L}[\sigma]$ the $T\bar T$-like deformed Liouville theory. It is given explicitly in \eqref{StildeLsigmat1} to first order, and we repeat it here with the $\bar \rho$-dependence (as discussed in \eqref{tcbarrho}) written out explicitly 
\begin{align}
	S_{\tilde{L}}=S_L- e^{-2\bar \rho}\frac{c}{48\pi} \int d^2x \, \frac{e^{-\sigma}}{16}\Bigg(&\frac{1}{2}(\partial \sigma)^4-2\hat{\Box}\sigma(\partial \sigma)^2\Bigg)+\mathcal{O}(e^{-4\bar \rho}). \label{StildeLsigmat1bis}
\end{align}	

Here all indices are raised and lowered with 
$\eta_\mn$, e.g. $(\partial \sigma)^2=\eta^{\mu\nu}\p_{\mu} \sigma \p_{\nu} \sigma$ and $\Box \sigma=\eta^{\mn}\partial_{\mu}\partial_{\nu}\sigma$. More generally, the ghost action should be included. 

In the traditional braneworld limit interpretation of \eqref{Zbw3d}, 
\ali{
	Z_{bw} \approx Z_{T\bar T}(\hat g)  \int \mathcal D \sigma \,  e^{i S_{\tilde L}[\sigma]|_{quadratic}} ,   \label{Zbwquadratic}
} 
the integral is interpreted to be over quadratic fluctuations around the saddle $\hat g$ satisfying the $\delta g$ EOM.  
As we show in Appendix \ref{AppTTbar}, the $T\bar{T}$-like deformed Liouville theory can be written in the form $S_{\tilde{L}}=S_L+\mathcal{O}(\sigma^3)$. Therefore at quadratic order in the small fluctuation $\sigma$ and general cut-off radius $\bar \rho$ 
we have 
\ali{
	S_{\tilde L}[\sigma]|_{quadratic}  =  \frac{c}{48\pi}\int d^2x\sqrt{-\hat{g}}\left( \sigma \hat{R}+\frac{1}{2}\hat{g}^{\mu\nu}\partial_{\mu}\sigma\partial_{\nu}\sigma \right) \label{SLquadratic} .
}
It is the timelike Liouville theory $S_L$ with $\lambda =0$ 
(whereas at higher orders, the Liouville theory will gain the discussed $T\bar T$-like corrections). 
The Liouville field takes on the interpretation of Weyl mode at the brane from a $T\bar T$ set free perspective. 
In a near-boundary expansion in $\epsilon^2 \equiv e^{-2 \bar \rho}$ or $(t \, c)$, it is related to the radion mode in braneworld language and asymptotic Weyl mode in the holographic integrated Weyl anomaly by   
\ali{
	\sigma = \phi + \mathcal O(\epsilon^2), \qquad \tilde \phi = \frac{\phi}{2} + \mathcal O(\epsilon^2)  \label{relationbetweenmodes}
}
so that asymptotically, the Liouville field in \eqref{Zbwquadratic} also takes on these alternative physical interpretations. 
This saddle-point result for $Z_{bw}$ reinterprets what is 
usually referred to as the ``cut-off CFT'' in holographic braneworld as the $T\bar T$-deformed CFT, and agrees asymptotically 
with the interpretation of the effective linearized gravity theory on the brane being given 
by timelike Liouville theory, as claimed for asymptotic (or near-boundary) branes in e.g. \cite{Compere:2008us,Suzuki:2022xwv}, but obtained here more generally in  \eqref{Zbwquadratic}.  
To summarize the saddle-point braneworld comparison, \eqref{Zbwquadratic} is the limiting, small fluctuation case of our more general  result \eqref{Zbw3d}. It is consistent asymptotically with previous work. 

Coming from the $T\bar T$ perspective, in section \ref{subsectTTSL}, 
we were able to obtain $S_{\tilde{L}}$ in terms of $\sigma$ only in a perturbative form. 
We did derive a non-perturbative expression for $S_{\tilde{L}}$ in \eqref{StildeLtildephiresult}, but in terms of the fluctuation $\tilde{\phi}$. 
A closed-form expression for $\tilde{\phi}(\sigma)$ would therefore allow us to write $S_{\tilde L}$ in terms of $\sigma$ also non-perturbatively.

\section{Braneworld} \label{sectbw}

For completeness, let us highlight in this section the connection between our notation and some of the language used in the original braneworld constructions of e.g. \cite{Randall:1999ee,Randall:1999vf,Karch:2000ct}. In particular, we want to address how the radion appears.

To construct a braneworld in AdS$_{D}$, one considers a fixed bulk geometry
\begin{equation}
	d\hat s^2=\hat G_{MN}(X)dX^MdX^N=d\rho^2+e^{2A(\rho)}\bar{g}(\rho,x)dx^2
\end{equation}
which satisfies the bulk equations of motion, and perturbs it to a new bulk geometry
\begin{equation}
	ds^{2}=G_{MN}(X)dX^MdX^N=d\rho^2+e^{2A(\rho)}\tilde{g}(\rho,x)dx^2
\end{equation}
by adding the linear perturbation $h(\rho,x)$ as follows
\begin{equation}
	\tilde{g}(\rho,x)=\bar{g}(\rho,x)+h(\rho,x).
\end{equation}
In standard braneworld constructions \cite{Randall:1999vf,Randall:1999ee,Karch:2000ct,Giddings:2000mu}, the brane is located at a general (not necessarily large) constant $\rho=\bar{\rho}$, while $h$ is infinitesimal (appearing linearly in EOMs and quadratically in the action), and generally one considers the original geometry to be vacuum AdS$_D$ in a maximally symmetric slicing, such that
\begin{equation}
	\bar{g}(\rho,x)=\hat g_{(0)}(x)
\end{equation}
with the particular form of the warping factor $A(\rho)$ depending on whether $\hat g_{(0)}(x)$ is flat, positively or negatively curved. The braneworld theory describes the gravitational dynamics induced at the boundary, with line element 
\begin{equation}
	ds^{2}\big|_{\rho=\bar{\rho}}=e^{2A(\bar{\rho})}\tilde{g}(\bar{\rho},x)dx^2=g(x)dx^2 
\end{equation}
where $g(x)$ is the induced metric. Then, as proposed by e.g.~\cite{Gubser:1999vj,Giddings:2000mu}, at the semi-classical level the braneworld system may be described by the following path integral  
\begin{equation}
	Z_{bw}=\int Dg \; Z_{tot}[g]= \int Dg \int_{G_{\partial}=g} D G e^{iS_{tot}[G]}.
\end{equation}
In the saddle-point limit of the bulk path integral, the linear perturbation solves the bulk equations of motion. For maximally symmetric slicings, 
the solution is given by \cite{Karch:2000ct}
\begin{equation}
	h^{\star}_{\mu\nu}(\rho,x)=h_{\mu\nu}^{TT}(\rho,x)-F(\rho)\nabla^{(0)}_{\mu}\nabla^{(0)}_{\nu}\phi(x)+\dot{A}g^{(0)}_{\mu\nu}\phi(x) \label{hbulkonshell}
\end{equation}
with $\dot{A}=\partial_{\rho}A$ and $\dot{F}=e^{-2A}$, in terms of the infinitesimal transverse-traceless modes $h_{\mn}^{TT}$ and the infinitesimal mode $\phi$.  
The field $\phi$ is called the radion mode in \cite{Karch:2000ct} (see also \cite{Charmousis:1999rg,Giannakis:2003hg}). 
Imposing the Neumann condition is equivalent to a semiclassical approximation of the integral over the induced metric, reducing further the expression for $h$. 
Setting $A(\rho)=\rho$ (note that $\lim\limits_{\rho\rightarrow \infty} \frac{A (\rho)}{\rho}=1$ for all warping factors of maximally symmetric slicings) and for a brane close to the asymptotic boundary, the perturbed bulk geometry takes the form 
\begin{equation}
	ds^{2}=G_{MN}(X)dX^MdX^N=d\rho^2+e^{2\rho}\left(g_{(0)}(x)+e^{-2\rho}g_{(2)}(x)+\mathcal{O}(e^{-4\rho})\right)dx^2.
\end{equation}
In our notation, the fluctuation is then  
\ali{
	h = (g_{(0)} - \hat g_{(0)}) + e^{-2\bar \rho} (g_{(2)} - \hat g_{(2)}) + e^{-4\bar \rho} (g_{(4)} - \hat g_{(4)}) + \cdots  
}
where we included again 
for generality the possibility of non-zero $\hat g_{(2),(4),...}$. 
In particular, in the $D=3$ case, there are no transverse traceless modes $h_{TT}$ left in \eqref{hbulkonshell} and the bulk on-shell expression for the fluctuation \eqref{hbulkonshell} is indeed consistent with the expressions for the metric components \eqref{g0phi} and \eqref{g2phi} given in our discussion of the BrH diffs in section \ref{subs3d}. 
That is, the Karch-Randall mode $\phi$ in \eqref{hbulkonshell} matches 
at the linear level with our asymptotic Weyl mode $\phi$. 
This is consistent with the argument in \cite{Karch:2000ct} that the scalar modes in $h^*$ can be gauged away in the $5D$ context precisely by a (infinitesimal) 
gauge-preserving diffeomorphism. 
In our work, in $3D$, we reinterpret this Karch-Randall radion field 
as related to the conformal factor $\sigma$ associated with the extension of the integrated Weyl anomaly into the bulk. 
This field is made dynamical through an artificial construction: it is first fixed by a Dirichlet boundary condition, and only after it is set free by hand. In the end it forms the inherently dynamical radion mode of the bulk gravity theory with Neumann boundary conditions.

Still in $D=3$, more recent work on braneworld holography \cite{Geng:2022slq,Geng:2022tfc} employs a strategy where the location of the brane is  
perturbed to $\rho = \bar \rho + \tilde \phi(x)$ with $\tilde \phi$ called the radion in those works. It matches our field $\tilde \phi$ in section \ref{subsectTTWeyl}.  
We have explained in detail in 
that section what the non-linear relation is between the diff mode $\phi$ and the radion $\tilde \phi$, as well as why they coincide at the linear level $\phi \approx 2\tilde \phi$. 

\section{Discussion and outlook} \label{sectdiscussion}

In this paper we have discussed the Neumann 
problem in $3D$ AdS-gravity by setting free the Dirichlet 
problem, $Z_{NBC} = \int Dg Z_{DBC}$ for $T=1$.  
This strategy 
allows to systematically discuss a holographic interpretation of braneworld theories by making use of the well-known holographic dualities for the Dirichlet case. 
Namely, it follows immediately from the holographic $T\bar T$ dictionary $Z_{DBC} = Z_{T\bar T}$ that the Neumann theory is dual to the  $T\bar T$ theory set free $\int Dg Z_{T\bar T}$. Compared to regular AdS/CFT, there is an additional parameter for measuring the location of the brane in the bulk  that is dual to the $T\bar T$ deformation parameter. 
Or in general dimensional language, the $T^2$ deformation. 
This provides a modern version of holographic braneworld statements, where now the unspecific ``cut-off CFT'' is understood to be a particular, well-defined deformation of the CFT.

Recent $3D/2d$ braneworld discussions, e.g.~\cite{Geng:2022slq,Geng:2022tfc,Deng:2022yll,Aguilar-Gutierrez:2023tic,Neuenfeld:2024gta}, derive effective gravity actions on the brane by integrating bulk gravity in Poincar\'e AdS up to a wiggly boundary, with the wiggles parametrized by the radial fluctuation 
$\tilde \phi(x)$, and treat that so-called radion $\tilde \phi(x)$ as the dynamical field in the dual $2d$ theory. We were interested in understanding this strategy from a holographic perspective. 
The holographic perspective makes clear that it is in fact the Weyl mode $\sigma$ of the induced metric that should be path integrated over\footnote{The Neumann boundary condition is semi-classically equivalent to setting free the induced metric (representing the Dirichlet-fixed boundary data from the gravitational perspective, or the field theory source from the boundary perspective). 
We should thus integrate over the induced metric's 
physical degree of freedom $\sigma$, which deviates from $\phi$ or $\tilde{\phi}$ as we flow into the bulk.}, and $\sigma$ is in general  different from $\tilde \phi$. It is also generally different from the Liouville field $\phi$, the 
	Weyl mode of the conformal boundary metric, that labels the Brown-Henneaux diffeomorphisms and describes the CFT's Weyl anomaly physics. We point out these differences and derive the holographic braneworld theory for a brane of tension $T=1$ at finite radial location to be given by  \eqref{Zbw3d}-\eqref{StildeLsigmat1bis}. The effective braneworld gravity $S_{\tilde L}$ 
is a $T\bar T$-like deformed timelike Liouville theory for $\sigma$, which can be constructed order by order in the deformation parameter. 
To quadratic order in fluctuations and for general cut-off radius it is simply a timelike Liouville theory with vanishing potential, but it receives corrections at higher orders. Our set free expressions for finite braneworld holography  
are more general than previous saddle-point braneworld ones, but reduce to them in the correct limits.

It was apparent in section \ref{subsectTTSL} that a significant simplification occurs when we consider only induced metrics $g = e^\sigma \hat g$ with zero curvature, $R=0$ or $\hat \Box \sigma = 0$.  
It suggests considering the separate theory $Z_{T\bar T}(\hat g) \int D \sigma \exp\{ i S_{\tilde F}[\sigma] \}$. It is unclear if it can be considered a well-defined sector of the braneworld theory $Z_{bw}$, and we leave investigation of this theory for future work.

Our procedure involved rewriting the $T\bar T$ theory in the form $Z_{T\bar T} = Z_{T\bar T}(\hat g) \exp\{ i S_{\tilde L}[\sigma] \}$ 
by dissecting  
the $3D$ AdS bulk theory (or DBC problem)  
using both bulk diffeomorphisms and gravitational Hamilton-Jacobi flows. We believe this dissection strategy to be fruitful. For example, as a side result, we found that the Liouville field $\phi$ not only describes the asymptotic conformal symmetry 
physics in AdS/CFT but also describes near-horizon physics of a conformally related bulk geometry. 
As another example, the obtained $S_{\tilde L}$ is also expected to be of separate use for investigating the conformal (CBC) problem, as it is related to the DBC problem by $Z_{CBC} = \int D \sqrt{-g} \, Z_{DBC} \exp\{ -\frac{i}{2\kappa} \int d^2 x \sqrt{-g} (K + 2) \}$.
   
A Liouville theory with a central charge that cancels the one of the CFT also appears in the non-critical string description of $T\bar T$ \cite{Callebaut:2019omt}. In the undeformed limit $t \ra 0$, that Liouville theory reduces to the timelike Liouville theory describing the holographic integrated Weyl anomaly, and thus to our $S_{\tilde L}$. 
We plan to further investigate the 
presented work in the context of the non-critical string and $2d$ gravity descriptions of $T\bar T$ \cite{Dubovsky:2017cnj,Dubovsky:2018bmo,Tolley:2019nmm,Cardy:2018sdv,Callebaut:2019omt,Ondo:2022zgf,Hirano:2025cjg}, as well as the mixed boundary condition proposal of \cite{Guica:2019nzm}, the conformal boundary condition problem \cite{Witten:2022xxp,Allameh:2025gsa,Anninos:2023epi,Coleman:2020jte} and other related works such as \cite{Kawamoto:2023wzj,Emparan:2022ijy,Emparan:2023dxm}. 
We are interested in extending our analysis beyond $T=1$ to discuss AdS$_2$ and dS$_2$ branes and 
associated braneworld constructions in AdS/bCFT, particularly in connection  to entanglement entropy and islands \cite{Donnelly:2018bef,Suzuki:2022xwv,Lewkowycz:2019xse,Hawking:2000da,Apolo:2023ckr,Deng:2023pjs,Afrasiar:2023nir}.

\acknowledgments{
	We thank Dionysios Anninos, Julian Arenz, Blanca Hergueta, Ruben Monten, Dominik Neuenfeld, Edgar Shaghoulian, Watse Sybesma and Konstantin Weisenberger for useful discussions. 
	We especially thank Dmitry Bagrets and Rodolfo Panerai for extensive and illuminating discussions on the work presented in this paper. The research of MS is funded by the Deutsche Forschungsgemeinschaft (DFG, German Research Foundation) – 
	Projektnummer 277101999 – TRR 183. 
}

	\newpage 
	
	\appendix 
	
	\section{Derivation of asymptotically AdS action with general boundary profile} \label{appendix:A}

	We present here the calculation leading to \eqref{SgravFGgeneral}, which appeared in the earlier work \cite{Carlip:2005tz} (see also \cite{Takayanagi:2018pml}). 
	Consider the following ansatz for a $3D$ bulk geometry $\mathcal{M}:FG(X)$
	\ali{
		FG(X): \; \; \; \; ds^2&=G_{MN}(X)dX^MdX^N=d\rho^2+\gamma_{\mu\nu}(X)dx^{\mu}dx^{\nu}
		\\
		&=d\rho^2+e^{2\rho}\left(g_{\mu\nu}^{(0)}(x)+e^{-2\rho}g^{(2)}_{\mu\nu}(x)+\dots\right)dx^{\mu}dx^{\nu},
	}
	with bulk coordinates $X:(\rho,x)$ and boundary coordinates $x$. The asymptotic boundary is reached in the limit $\rho \rightarrow \infty$. However, we let the timelike outer boundary $\partial \mathcal{M}$ be located at
	\ali{
		\partial \mathcal{M}: \; \; \; \; \; \; \; \; \; \; \; \; \rho=F(x).
	} 
	The action for $3D$ Einstein gravity with negative cosmological constant, supplemented by the Gibbons-Hawking boundary term and holographic counterterm and evaluated on the manifold $\mathcal{M}$ is given by 
	\ali{S_{grav}[FG(\rho < F(x),x)]=\frac{1}{2\kappa} \int_{\mathcal M} 
		d^{3} X \sqrt{|G|} \left( R_G  + 2  \right) + 
		\frac{1}{\kappa} \int_{\p \mathcal M} 
		d^2 x \sqrt{|g|} \, \left(K-1\right),
	}
	where the induced metric at the outer boundary is
	\begin{equation}
		g_{\mu\nu}(x)dx^{\mu}dx^{\nu}=ds^2\Big|_{\rho=F(x)}=\big(\gamma_{\mu\nu}(X)+\partial_{\mu} F(x)\partial_{\nu} F(x)\big)\Big|_{\rho=F(x)}dx^{\mu}dx^{\nu}.
	\end{equation}
	Using that for $2 \times 2$ matrices $A,B$ we have the following properties of determinants $det(A+B)=(det A) ( 1+Tr(A^{-1}B))+det B$ and $det(\lambda A)=\lambda^2 det A$, we find
	\begin{equation}
		\sqrt{|G|}=\sqrt{|\gamma|}=e^{2\rho}\sqrt{|g_{(0)}|}+\frac{1}{2}\sqrt{|g_{(0)}|} \, g_{(0)}^{\mu\nu}g^{(2)}_{\mu\nu}+\mathcal{O}(e^{-2\rho})
	\end{equation}
	and
	\begin{align}
		\sqrt{|g|}&=\left(\sqrt{| \gamma|}\sqrt{1+\gamma^{\mu\nu}\partial_{\mu} F\partial_{\nu} F}\right)\Big|_{\rho=F}
		\\
		&=\sqrt{|g_{(0)}|}\left(e^{2F}+\frac{1}{2}g_{(0)}^{\mu\nu}g^{(2)}_{\mu\nu}+\frac{1}{2}g^{\mu\nu}_{(0)}\partial_{\mu} F\partial_{\nu} F\right)+\mathcal{O}(e^{-2F}) . 
	\end{align}
	The components of the normal $n_{M}=\frac{\partial_{M}\left(\rho-F\right)}{\sqrt{|G^{AB}\partial_A (\rho-F)\partial_B (\rho-F)|}}$ are given by
	\begin{equation}
		n_{\rho}=\frac{1}{\sqrt{1+\gamma^{\alpha\beta}\partial_{\alpha} F\partial_{\beta} F}}, \;\;\;\; n_{\mu}=-\frac{\partial_{\mu} F}{\sqrt{1+\gamma^{\alpha\beta}\partial_{\alpha} F\partial_{\beta} F}}.
	\end{equation}
	The trace of the extrinsic curvature on the boundary is thus
	\begin{equation}
		K=\left(G^{MN}\nabla^{G}_{M}n_{N}\right)\Big|_{\rho=F}=2-e^{-2F}g_{(0)}^{\mu\nu}g^{(2)}_{\mu\nu}-e^{-2F}g^{\mu\nu}_{(0)}\nabla_{\mu}^{(0)}\partial_{\nu} F+\mathcal{O}(e^{-4F}),
	\end{equation}
	where we used $\Gamma^{\rho}_{\rho\rho}=\Gamma^{\mu}_{\rho\rho}=0$, $\Gamma^{\rho}_{\mu\nu}=-\frac{1}{2}\partial_{\rho}\gamma_{\mu\nu}$ and $\gamma^{\mu\nu}=e^{-2\rho}\left(g_{(0)}^{\mu\nu}-e^{-2\rho}g^{\mu\nu}_{(2)}+\mathcal{O}(e^{-4\rho})\right)$. Therefore, we have
	\begin{equation}
		\sqrt{|g|}K=\sqrt{|g_{(0)}|}\left(2 e^{2F}+g^{\mu\nu}_{(0)}\partial_{\mu} F\partial_{\nu} F\right)-\partial_{\mu}\left(\sqrt{|g_{(0)}|}g^{\mu\nu}_{(0)}\partial_{\nu} F\right)+\mathcal{O}(e^{-2F}). \label{gammaKCarlip}
	\end{equation}
	Using that $R_G=-6$ when the Einstein equations are satisfied, we have that the bulk and boundary terms evaluate respectively to
	\begin{align}
		\frac{1}{2 \kappa}\int_{\mathcal{M}} d^3X \sqrt{|G|}\left(R_G+2\right)&=-\frac{2}{\kappa}\int d^2 x \sqrt{|g_{(0)}|} \int\limits^{\rho=F}d\rho\left(e^{2\rho}+\frac{1}{2}g_{(0)}^{\mu\nu}g^{(2)}_{\mu\nu}+\mathcal{O}(e^{-2\rho})\right) 
		\\
		&=-\frac{1}{\kappa}\int d^2 x \sqrt{|g_{(0)}|}\left(e^{2F}+Fg_{(0)}^{\mu\nu}g^{(2)}_{\mu\nu}+\mathcal{O}(e^{-2F})\right)+\cdots,
	\end{align} 
	and
	\begin{align}
		\frac{1}{\kappa}\int_{\partial \mathcal{M}} d^2x \sqrt{|g|}\left(K-1\right)=\frac{1}{\kappa}\int d^2x \sqrt{|g_{(0)}|}\Big(e^{2F}&+\frac{1}{2}g_{(0)}^{\mu\nu}\partial_{\mu} F\partial_{\nu} F \nonumber
		\\
		&-\frac{1}{2}g_{(0)}^{\mu\nu}g^{(2)}_{\mu\nu}+\mathcal{O}(e^{-2F})\Big),
	\end{align}
	where we denoted with the dots the contributions from the lower bound of the radial integral in the bulk term and we assumed the vanishing of the total derivative term in \eqref{gammaKCarlip}. Therefore, we obtain
	\ali{
		& S_{grav}[FG(\rho < F(x), x)] \nonumber \\
		&\quad =  \frac{1}{2\kappa}\int d^2x \sqrt{|g_{(0)}|}\left(g_{(0)}^{\mn}\partial_\mu F\partial_\nu F - g_{(0)}^{\mn} g^{(2)}_{\mn}\left(1+2 F\right)+\mathcal{O}(e^{-2F})\right)+\cdots . 
	}

	\section{BTZ and Liouville}  \label{appendix:B}
	
	Consider the non-rotating $BTZ$ black hole solution in FG coordinates (Banados form)
	\begin{equation}
		ds^{2}_{BTZ}=G_{MN}(X)dX^{M}dX^{N}=d\rho^2-2e^{2\rho}dzd\bar{z}+Ldz^2+\bar{L}d\bar{z}^2-\frac{e^{-2\rho}}{2}L\bar{L}dzd\bar{z}.
	\end{equation}
	The parameters $L=\bar{L}$ are related to the mass $M$ of the black hole as
	\begin{equation}
		L=\bar{L}=\frac{\kappa}{2\pi}M.
	\end{equation}
	The black hole horizon is located at $\rho=\rho_+$, with
	\begin{equation}
		\rho_+=\frac{1}{2}\log\left(\frac{\kappa}{4\pi}M\right).
	\end{equation}
	\par We want to evaluate the action
	\ali{S_{grav}[FG(\rho < \bar{\rho},x)]=\frac{1}{2\kappa} \int_{\mathcal M} 
		d^{3} X \sqrt{|G|} \left( R_G  + 2  \right) 
		+\frac{1}{\kappa} \int_{\p \mathcal M} 
		d^2 x \sqrt{|g|} \, \left(K-1\right) 
	}
	on the non-rotating BTZ solution $FG:\mathcal{M} $ with boundary $\partial \mathcal{M}$ located at $\rho=\bar{\rho}$ and $\rho=\rho^+$. Starting with the bulk term and using $R_G=-6$, we find
	\begin{align}
		\frac{1}{2\kappa} \int_{\mathcal{M}} 
		d^{3} X \sqrt{|G|} \left( R_G  + 2  \right)=&\frac{2}{\kappa}\int dz d\bar{z}\int^{\bar{\rho}}_{\rho_+} \left(\frac{e^{-2\rho}}{4}L^2-e^{2\rho}\right)d\rho
		\\
		=&\frac{1}{\kappa}\int dz d\bar{z} \left(L-e^{2\bar{\rho}}-\frac{e^{-2\bar{\rho}}}{4}L^2\right) \, .
	\end{align}
	The trace of the extrinsic curvature at the outer boundary is given by
	\begin{equation}
		K=2\frac{\left(4+e^{-4\bar{\rho}} L^2\right)}{\left(4-e^{-4\bar{\rho}} L^2\right)},
	\end{equation}
	while at the inner boundary it differs by an overall sign (and with $\bar \rho \ra \rho_+$). Moreover, the area element at the outer boundary is
	\begin{equation}
		\int_{\partial \mathcal{M}} d^2x \sqrt{|g|}=\int dz d\bar{z}\left(e^{2\bar{\rho}}-\frac{e^{-2\bar{\rho}}}{4}L^2\right),
	\end{equation}
    and analogously at the inner boundary (with $\bar \rho \ra \rho_+$). Putting everything together, we obtain 
	\begin{equation} \label{SgravBTZ}
		S_{grav}[FG(\rho < \bar{\rho},x)]=-\frac{1}{\kappa}\int dz d\bar{z} \left(L+\frac{e^{-2\bar{\rho}}}{2}L^2\right) \, .
	\end{equation}
	Therefore, in the asymptotic limit we find
	\begin{equation}
		\lim\limits_{\bar{\rho} \rightarrow \infty}	S_{grav}[FG(\rho < \bar{\rho},x)]=-\frac{M}{2\pi}\int dz d\bar{z} \, .
	\end{equation}
	Note that the finite contribution to the asymptotic value of the on-shell action originates entirely from the combination of the lower bound of the radial integration in the bulk term and the GHY term at the inner boundary, i.e.~from the horizon. Had we excluded the GHY boundary term contribution at the horizon, we would have obtained an opposite overall sign.
	\par Introducing the functions
	\begin{equation}
		f(z)=e^{\sqrt{\frac{2\kappa}{\pi} M}\, z}, \; \; \; \; \; \bar{f}(\bar{z})=e^{\sqrt{\frac{2\kappa}{\pi} M} \, \bar{z}}
	\end{equation}
	such that
	\begin{equation}
		L=-\frac{1}{2}\{f,z\}=\frac{\kappa}{2\pi} M=-\frac{1}{2}\{\bar{f},\bar{z}\}=\bar{L},
	\end{equation} 
	and 
	\begin{equation}
		-2dzd\bar{z}=-2e^{\phi}dfd\bar{f}=e^{\phi}\hat{g}^{(0)}_{\mu\nu}dx^{\mu}dx^{\nu},
	\end{equation}
	with 
	\begin{equation}
		\phi=\log\left(\frac{\partial z}{\partial f}\frac{\partial \bar{z}}{\partial \bar{f}}\right)=-\log\left(\frac{2 \kappa}{\pi} M f \bar{f}\right)
	\end{equation}
	we can write the line element for $\mathcal{M}$ as
	\begin{equation}
		ds^{2}_{BTZ}=G_{MN}(X)dX^{M}dX^{N}=d\rho^2-2e^{2\rho}e^{\phi}dfd\bar{f}+\frac{L}{f^{\prime 2}}df^2+\frac{\bar{L}}{\bar{f}^{\prime 2}}d\bar{f}^2-\frac{e^{-2\rho}}{2}e^{\phi}L \bar{L}dfd\bar{f}, 
	\end{equation}
	with $f^{\prime}=\frac{\partial f}{\partial z}$ and $\bar{f}^{\prime}=\frac{\partial \bar{f}}{\partial \bar{z}}$. The non-rotating BTZ solution is related by the Brown-Henneaux diffeomorphisms to the Poincaré solution $\widehat{FG}$ with line element
	\begin{equation}
		ds^{2}_{Poincare}=\hat{G}_{MN}(X)dX^M dX^N=d\rho^2-2e^{2\rho}dfd\bar{f}.
	\end{equation}
	Sending $(z,\bar{z})\rightarrow (f,\bar{f})$ and $L,\bar{L}\rightarrow0$ in \eqref{SgravBTZ}, it is easy to see that the asymptotic value of the on-shell action evaluated on the Poincaré solution vanishes
	\begin{equation}
		\lim\limits_{\bar{\rho} \rightarrow \infty}	S_{grav}[\widehat{FG}(\rho < \bar{\rho},x)]=0 \, .
	\end{equation}
	Therefore, we expect the difference
	\begin{equation}
		\lim\limits_{\bar{\rho} \rightarrow \infty} \left(	S_{grav}[FG(\rho < \bar{\rho},x)]-S_{grav}[\widehat{FG}(\rho < \bar{\rho},x)]\right)=-\frac{M}{2\pi}\int dz d\bar{z}
	\end{equation}
	originating at the BTZ horizon to be entirely captured by the Liouville action  
	\begin{equation}
		S_L=\frac{c}{48\pi}\int d^2x\sqrt{|\hat{g}_{(0)}|}\left(\phi \hat{R}_{(0)}+\frac{1}{2}\hat{g}_{(0)}^{\mu\nu}\partial_{\mu} \phi\partial_{\nu} \phi\right).
	\end{equation}
	Using that $\hat{R}_{(0)}=0$,
	\begin{equation}
		\frac{1}{2}\hat{g}_{(0)}^{\mu\nu}\partial_{\mu} \phi\partial_{\nu} \phi=-\frac{1}{f\bar{f}}
	\end{equation}
	and 
	\begin{align}
		\int d^2x\sqrt{|\hat{g}^{(0)}|}=\int df d\bar{f}=\frac{2\kappa}{\pi} M \int dzd\bar{z} f\bar{f},
	\end{align}
	we find that indeed
	\begin{equation}
		S_L=\lim\limits_{\bar{\rho} \rightarrow \infty} \left(	S_{grav}[FG(\rho < \bar{\rho},x)]-S_{grav}[\widehat{FG}(\rho < \bar{\rho},x)]\right).
	\end{equation}

	\section{Derivation of Banados action with general boundary profile} \label{appendix:D}
	
	Let the $3D$ manifold $\mathcal{M}$ be given by the Banados solution
	\begin{align}
		Ban(X): \; \; \; \; \; ds^2_{Banados}&=G_{MN}(X)dX^MdX^N
		\\
		&=d\rho^2+\gamma_{\mu\nu}(X)dx^{\mu}dx^{\nu}
		\\
		&=d\rho^2+e^{2\rho}\left(g^{(0)}_{\mu\nu}(x)+e^{-2\rho}g^{(2)}_{\mu\nu}(x)+e^{-4\rho}g^{(4)}_{\mu\nu}(x)\right)dx^{\mu}dx^{\nu},
	\end{align}
	with $g^{(0)}_{\mu\nu}(x)dx^{\mu}dx^{\nu}=-2dzd\bar{z}$, $g^{(2)}_{\mu\nu}(x)dx^{\mu}dx^{\nu}=Ldz^2+\bar{L}d\bar{z}^2$, $g^{(4)}_{\mu\nu}(x)dx^{\mu}dx^{\nu}=-\frac{L \bar{L}}{2}dzd\bar{z}$ and with a timelike outer boundary $\partial \mathcal{M}$ located at
	\ali{
		\partial \mathcal{M}: \; \; \; \; \; \; \; \; \; \; \; \; \rho=F(x).
	} 
	Setting $L=\bar{L}=0$, hence $g_{(2)}=g_{(4)}=0$, the above Banados solution reduces to Poincaré AdS$_3$, which we denote as $Poinc(X)$. The induced line element on $\partial \mathcal{M}$ is
	\begin{equation}
		g_{\mu\nu}(x)dx^{\mu}dx^{\nu}=ds_{Banados}^2\Big|_{\rho=F(x)}=\left(\gamma_{\mu\nu}(X)+\partial_{\mu} F(x)\partial_{\nu} F(x)\right)\Big|_{\rho=F(x)}dx^{\mu}dx^{\nu} . 
	\end{equation}
	The square roots of the absolute values of the determinants of the bulk metric $G_{MN}$, of the 2d metric $\gamma_{\mu\nu}$ and of the induced metric $g_{\mu\nu}$ are given by
	\begin{align}
		&\sqrt{|G|}=\sqrt{|\gamma|}=e^{2\rho}\left(1-e^{-4\rho}\frac{L\bar{L}}{4}\right), 
		\\
		&\sqrt{|g|}= e^{2F}\left(1-e^{-4F}\frac{L\bar{L}}{4}\right) \sqrt{1+(\p F)^2 }\Big|_{\rho=F},
	\end{align}
	with $(\p F)^2 \equiv \gamma^\mn \p_\mu F \p_\nu F$. The components of the normal are given by $n_{\rho}=(1+(\p F)^2)^{-1/2}$ and $ n_{\mu}=-\partial_{\mu} F(1+(\p F)^2)^{-1/2}$. The trace of the extrinsic curvature $K=\left(G^{MN}\nabla^{G}_{M}n_{N}\right)\Big|_{\rho=F}$ on the boundary is 
	\begin{equation}
		K=\left[-\frac{\partial_{\mu} F\partial_{\nu} F\partial_{\rho}\gamma^{\mu\nu}}{2\left(1+(\p F)^2 \right)^{3/2}}+\gamma^{\mu\nu}\nabla^{(\gamma)}_{\mu}n_{\nu}+\frac{2 (1+e^{-4\rho}\frac{L\bar{L}}{4})(1-e^{-4\rho}\frac{L\bar{L}}{4})^{-1}}{\sqrt{1+(\p F)^2}}\right]\Bigg|_{\rho=F},
	\end{equation}
	where $\nabla_{\mu}^{(\gamma)}$ denotes covariant differentiation with respect to the $2d$ metric $\gamma_{\mu\nu}$ and we used $\Gamma^{\rho}_{\rho\rho}=\Gamma^{\mu}_{\rho\rho}=0$, $\Gamma^{\rho}_{\mu\nu}=-\frac{1}{2}\partial_{\rho}\gamma_{\mu\nu}$ and $\frac{1}{2}\gamma^{\mu\nu}\partial_{\rho}\gamma_{\mu\nu}=2 (1+e^{-4\rho}\frac{L\bar{L}}{4})(1-e^{-4\rho}\frac{L\bar{L}}{4})^{-1}$. Therefore, we have
	\begin{align}
		\sqrt{|g|}K=&2e^{2F}\left(1+e^{-4F}\frac{L\bar{L}}{4}\right)-e^{2F}\left(1-e^{-4 F}\frac{L\bar{L}}{4}\right)\Bigg[\frac{\partial_{\mu} F\partial_{\nu} F\partial_{\rho}\gamma^{\mu\nu}}{2\left(1+(\p F)^2\right)} \nonumber
		\\
		&+\sqrt{1+(\p F)^2} \gamma^{\mu\nu}\nabla^{(\gamma)}_{\mu}\left(\frac{\partial_{\nu} F}{\sqrt{1+(\p F)^2}}\right)\Bigg]\Bigg|_{\rho=F}.
	\end{align}
	We want to evaluate the action for $3D$ Einstein gravity with negative cosmological constant, supplemented by the Gibbons-Hawking boundary term and tension term on the manifold $\mathcal{M}$  
	\ali{S_{tot}[Ban(\rho < F(x),x)]=\frac{1}{2\kappa} \int_{\mathcal M} 
		d^{3} X \sqrt{|G|} \left( R_G  + 2  \right) 
		+\frac{1}{\kappa} \int_{\p \mathcal M} 
		d^2 x \sqrt{|g|} \, \left(K-T\right).
	}
	
	Using $R_{(0)}=0$ and thus $R_G=-6=e^{-2\rho}R_{(0)}-6$ \footnote{This artificially introduces back $R_{(0)}$ in the action expressions below, to 
		illustrate where the $\phi R_{(0)}$ terms in the Liouville actions in the main text come from.}, we have that the bulk and boundary terms evaluate respectively to
	\begin{align}
		\frac{1}{2\kappa}\int_{\mathcal{M}} d^3X \sqrt{|G|}\left(R_G+2\right)=\frac{1}{2\kappa}\int_{\partial \mathcal{M}} d^2 x \sqrt{|g_{(0)}|}\Bigg[ &  R_{(0)}\left(F+e^{-4F}\frac{L\bar{L}}{16}\right) \nonumber
		\\
		&-2e^{2F}-e^{-2F}\frac{L\bar{L}}{2}\Bigg]+\cdots,
	\end{align} 
	and
	\begin{align}
		\frac{1}{\kappa}\int_{\partial \mathcal{M}} d^2x \sqrt{|g|}&\left(K-T\right)=\frac{1}{\kappa}\int_{\partial \mathcal{M}} d^2 x \sqrt{|g_{(0)}|} \Bigg\{2e^{2F}\left(1+e^{-4F}\frac{L\bar{L}}{4}\right) \nonumber
		\\
		&-e^{2F}\left(1-e^{-4 F}\frac{L\bar{L}}{4}\right)\Bigg[\frac{\partial_{\mu} F\partial_{\nu} F\partial_{\rho}\gamma^{\mu\nu}}{2\left(1+(\p F)^2\right)}+T\sqrt{1+(\p F)^2}  \nonumber
		\\
		&+\sqrt{1+(\p F)^2} \gamma^{\mu\nu}\nabla^{(\gamma)}_{\mu}\left(\frac{\partial_{\nu} F}{\sqrt{1+(\p F)^2}}\right)\Bigg]\Bigg|_{\rho=F}\Bigg\} .
	\end{align}
	Here and in the following, we denote with the dots the contributions from lower bound of radial integration in the Einstein-Hilbert term. Therefore, we find 
	\begin{align}
		S_{tot}[Ban(\rho &< F(x),x)]=\frac{1}{2\kappa}\int_{\partial \mathcal{M}} d^2 x \sqrt{|g_{(0)}|}\Bigg\{   R_{(0)}\left(F+e^{-4F}\frac{L\bar{L}}{16}\right) \nonumber
		\\
		&+ 2e^{2F}\left(1+e^{-4F}\frac{L\bar{L}}{4}\right)-e^{2F}\left(1-e^{-4 F}\frac{L\bar{L}}{4}\right)\Bigg[\frac{\partial_{\mu} F\partial_{\nu} F\partial_{\rho}\gamma^{\mu\nu}}{1+(\p F)^2} \nonumber
		\\
		&+2\sqrt{1+(\p F)^2}\left(T+\gamma^{\mu\nu}\nabla^{(\gamma)}_{\mu}\left(\frac{\partial_{\nu} F}{\sqrt{1+(\p F)^2}}\right)\right) \Bigg]\Bigg|_{\rho=F}\Bigg\}+\cdots.
	\end{align}
	Here $(\p F)^2=\gamma^{\mu\nu}\partial_{\mu}F\partial_{\nu}F$, and we repeat for reference that $\nabla_{\mu}^{(\gamma)}$ denotes covariant differentiation with respect to the $2d$ metric $\gamma_{\mu\nu}$, and the dots denote potential contributions from the lower bound of radial integration in the EH term. Note that the notation $\big|_{\rho=F}$ after the brackets means evaluated at $\rho=F$ \textit{after} derivatives are taken. Let's now look at some interesting cases. Fixing the radial location of the boundary to a constant $\rho=\bar{\rho}$ value, we have
	\begin{align}
		S_{tot}[Ban(\rho < \bar{\rho},x)]=&\frac{1}{2\kappa}\int_{\partial \mathcal{M}} d^2 x \sqrt{|g_{(0)}|}\Bigg[   R_{(0)}\left(\bar{\rho}+e^{-4\bar{\rho}}\frac{L\bar{L}}{16}\right) \nonumber
		\\
		&+ 2e^{2\bar{\rho}}\left(1+e^{-4\bar{\rho}}\frac{L\bar{L}}{4}\right)-2Te^{2\bar{\rho}}\left(1-e^{-4 \bar{\rho}}\frac{L\bar{L}}{4}\right)\Bigg]+\cdots,
	\end{align}
	which reduces to the following when the tension term is fixed to the  counterterm ($T=1$)
	\begin{equation}
		S_{grav}[Ban(\rho < \bar{\rho},x)]=\frac{1}{2\kappa}\int_{\partial \mathcal{M}} d^2x \sqrt{|g_{(0)}|} \left( R_{(0)}\left(\bar{\rho}+e^{-4\bar{\rho}}\frac{L\bar{L}}{16}\right)+e^{-2\bar{\rho}}L\bar{L}\right)+\cdots.
	\end{equation}
	In particular, for the vacuum case $L=\bar{L}=0$ we obtain
	\begin{equation} \label{IgravbanadosemptyT1rhocst}
		S_{grav}[Poinc(\rho < \bar{\rho},x)]=\frac{\bar{\rho} - \rho_+}{2\kappa}\int_{\partial \mathcal{M}} d^2x \sqrt{|g_{(0)}|}  R_{(0)}.
	\end{equation}
	Note in this case that additional contributions from the inner boundary at $\rho=\rho_+=-\infty$, the Poincaré horizon, vanish. 
	For general $L$ and $\bar{L}$, we can set  $F(x)=\bar{\rho}+\frac{1}{2}\phi(x^a)+\mathcal{O}(e^{-2\bar{\rho}})$ while fixing the tension term to the renormalizing counterterm ($T=1$), to recover
	\begin{align}
		S_{grav}[Ban&(\rho < \bar{\rho}+\phi/2+\mathcal{O}(e^{-2\bar{\rho}}),x)]=S_{grav}[Ban(\rho < \bar{\rho},x)] \nonumber
		\\
		&+\frac{c}{48\pi}\int_{\partial \mathcal{M}} d^2x \sqrt{|g_{(0)}|}\Big(\phi R_{(0)}+\frac{1}{2}g^{\mu\nu}_{(0)}\partial_{\mu} \phi \partial_{\nu} \phi  \Big)+\mathcal{O}(e^{-2\bar{\rho}}). 
	\end{align}
	Let us now focus on Poincaré AdS$_3$ (i.e. $L=\bar{L}=0$), for which we have
	\begin{align}
		S_{tot}[Poinc(\rho &< F(x),x)]=\frac{1}{2\kappa}\int_{\partial \mathcal{M}} d^2 x \sqrt{|g_{(0)}|}\Bigg\{  (F-\rho_+) R_{(0)}+ 2e^{2F}-e^{2F}\Bigg[-\frac{2(\p F)^2}{1+(\p F)^2}\nonumber
		\\
		&+2\sqrt{1+(\p F)^2}\left(T+\gamma^{\mu\nu}\nabla^{(\gamma)}_{\mu}\left(\frac{\partial_{\nu} F}{\sqrt{1+(\p F)^2}}\right)\right) \Bigg]\Bigg|_{\rho=F}\Bigg\}.
	\end{align}
	By setting $F(x)=\bar{\rho}+\tilde{\phi}(x)$, we can re-express this action in terms of the fluctuation $\tilde{\phi}(x)$ around the constant radial location $\rho=\bar{\rho}$
	\begin{align}
		S_{tot}[Poinc(\rho &< \bar{\rho}+\tilde{\phi}(x),x)]=\frac{1}{2\kappa}\int d^2x \sqrt{|g_{(0)}|}\Bigg\{ -\log \epsilon \; R_{(0)}-\rho_+R_{(0)}+ \tilde{\phi}R_{(0)} \nonumber
		\\
		&+\frac{2e^{2\tilde{\phi}}}{\epsilon^2}+2(\partial \tilde{\phi})^2-2g^{\mu\nu}_{(0)}\partial_{\mu}\partial_{\nu} \tilde{\phi} -\frac{2e^{2\tilde{\phi}}}{\epsilon^2}T\sqrt{1+\epsilon^2 e^{-2\tilde{\phi}}(\partial \tilde{\phi})^2} \nonumber
		\\
		&+\frac{g^{(0)\mu\nu}\partial_{\mu} \tilde{\phi}\left[-2\epsilon^2e^{-2\tilde{\phi}}\partial_{\nu}\tilde{\phi}(\partial \tilde{\phi})^2+\epsilon^2 e^{-2\tilde{\phi}}\partial_{\nu}(\partial \tilde{\phi})^2\right]}{1+\epsilon^2 e^{-2\tilde{\phi}}(\partial \tilde{\phi})^2}\Bigg\}, \label{IgravbanadosemptyT1phi}
	\end{align}
	with $(\partial \tilde{\phi})^2=g_{(0)}^{\mu\nu}\partial_{\mu}\tilde{\phi}\partial_{\nu}\tilde{\phi}$ and $\epsilon=e^{-\bar{\rho}}$. Expanding for infinitesimal $\tilde{\phi}$ and assuming vanishing of the total derivative, we find 
	\begin{align}
		&S_{tot}[Poinc(\rho < \bar{\rho}+\tilde{\phi}(x),x)]= \nonumber
		\\
		&\frac{c}{24\pi}\int d^2x \sqrt{|g_{(0)}|}\Big\{ (\bar{\rho}-\rho_++\tilde{\phi})R_{(0)}+\left(2-T\right)(\partial \tilde{\phi})^2+2(1-T)e^{2\bar{\rho}}e^{2\tilde{\phi}}\Big\}+\mathcal{O}(\tilde{\phi}^3).
	\end{align}
	The action $S_{\tilde{L}}$, defined as
	\begin{equation}
		S_{\tilde{L}}=S_{grav}[Poinc(\rho < \bar{\rho}+\tilde{\phi}(x),x)]-S_{grav}[Poinc(\rho < \bar{\rho},x)],
	\end{equation}
	takes the following exact form
	\begin{align}
		S_{\tilde{L}}=\frac{1}{2\kappa}\int d^2x  \sqrt{|g_{(0)}|}\Bigg\{&\tilde{\phi}R_{(0)}+\frac{2e^{2\tilde{\phi}}}{\epsilon^2}+2(\partial \tilde{\phi})^2
		-\frac{2e^{2\tilde{\phi}}}{\epsilon^2}\sqrt{1+\epsilon^2 e^{-2\tilde{\phi}}(\partial \tilde{\phi})^2}
		\nonumber
		\\
		&+\frac{\partial^{\mu} \tilde{\phi}\left[-2\epsilon^2e^{-2\tilde{\phi}}\partial_{\mu}\tilde{\phi}(\partial \tilde{\phi})^2+\epsilon^2 e^{-2\tilde{\phi}}\partial_{\mu}(\partial \tilde{\phi})^2\right]}{1+\epsilon^2 e^{-2\tilde{\phi}}(\partial \tilde{\phi})^2}\Bigg\},
	\end{align}
	with $\partial^{\mu}\tilde{\phi}=g^{\mu\nu}_{(0)}\partial_{\nu}\tilde{\phi}$. This result can be written perturbatively in an asymptotic $\epsilon\rightarrow 0$ ($\bar{\rho}\rightarrow \infty$) expansion. The first order result is
	\begin{equation}
		S_{\tilde{L}}=\frac{1}{2\kappa}\int d^2x\sqrt{|g_{(0)}|}\Big\{ \tilde{\phi}R_{(0)}+(\p \tilde{\phi})^2
		+ \frac{\epsilon^2}{e^{2\tilde{\phi}}}\left(\frac{1}{4}(\p \tilde{\phi})^4-\partial_{\mu}\partial^{\mu} \tilde{\phi}(\p \tilde{\phi})^2\right)
		+\mathcal{O}(\epsilon^4)\Big\} . 
	\end{equation}
	Note that we dropped total derivatives and integrated by parts. In terms of the conformal factor $\phi$ labelling the Brown-Henneaux diffeomorphisms, it takes the form
	\begin{equation}
		S_{\tilde{L}}=\frac{c}{48\pi}\int d^2x\sqrt{|g_{(0)}|}\Bigg\{ \phi R_{(0)}+\frac{1}{2}(\p \phi)^2 
		-\frac{\epsilon^2}{8e^{\phi}}\Bigg[\frac{3}{4}(\p \phi)^4-\partial^{\mu} \phi\partial_{\mu}(\p \phi)^2+R_{(0)}(\p \phi)^2  
		\Bigg]+\mathcal{O}(\epsilon^4)\Bigg\}, 
	\end{equation}
	with $(\p \phi)^2=g_{(0)}^{\mu\nu}\partial_{\mu}\phi\partial_{\nu}\phi$. Integrating by parts, denoting explicitly $g_{\mu\nu}^{(0)}=\eta_{\mu\nu}$ and $R_{(0)}=0$ this reduces to
	\begin{align}
		S_{\tilde{L}}=&\frac{c}{48\pi}\int d^2x\Bigg\{\frac{1}{2}(\p \phi)^2 +\frac{\epsilon^2}{32}e^{-\phi}(\p \phi)^4-\frac{\epsilon^2}{8}e^{-\phi}(\p\phi)^2\partial_{\mu}\partial^{\mu}\phi+\mathcal{O}(\epsilon^4)\Bigg\}.
	\end{align}

\section{Solution of the $T\bar T$-like flow equation} \label{AppTTbar}
We provide here the essential steps in the derivation of the $T\bar T$-like deformed Liouville theory 
\ali{
	S_{\tilde L} \equiv S_{\tilde L}^{(t)} 
}
defined as the solution to the trace flow equation
\ali{
	t_{\tilde L \, \mu}^\mu=\frac{c}{12}\hat \Box \sigma + t \, e^{-\sigma} \mathcal O^{\tilde L}_{T\bar T,\hat g}, \qquad \mathcal O^{\tilde L}_{T\bar T,\hat g} \equiv t_{\tilde L}^\mn t_{\tilde L}^\ab \hat g_{\alpha\mu} \hat g_{\beta\nu} - (t_{\tilde L}^\mn \hat g_\mn)^2 . \label{traceflowliouapp} 
}
with stress tensor $t^{\tilde L}_\mn = \frac{4\pi}{\sqrt{\hat g}} \frac{\delta S_{\tilde L}}{\delta \hat g^\mn}$, which we claim to be  equivalent to the action flow
\ali{
	\frac{d}{dt} S_{\tilde L}^{(t)} &= \frac{1}{4\pi} \int d^2 x \sqrt{-\hat g} \,  e^{-\sigma} \mathcal O^{\tilde L}_{T\bar T,\hat g} \label{actionflowapp} 
}
with seed action given by the $\lambda=0$ timelike Liouville theory for the Liouville field $\sigma$
\ali{
	S_{\tilde L}^{(0)} &=S_L= \frac{c}{48\pi} \int d^2 x \sqrt{-\hat g} \left(\sigma \hat R + \frac{1}{2} \hat g^\mn \p_\mu \sigma \p_\nu \sigma \right)  
}   
whose stress tensor is
\ali{
	t_\mn^{\tilde L (0)} = 
	-\frac{c}{24} \left(-\p_\mu \sigma \p_\nu \sigma + \frac{1}{2} \hat g_\mn   \hat g^\ab \p_\al \sigma \p_\beta \sigma   + 2 (\p_\mu \p_\nu \sigma - \hat g_\mn \hat \Box \phi ) \right). \label{tLapp} 
}
Throughout this appendix indices are raised and lowered with $\hat g_\mn$, which is fixed to be flat. To begin, we write the deformed action, Lagrangian and Lagrangian density as expansions in the deformation parameter $t$
\begin{equation}
	S^{(t)}_{\tilde{L}}=\int d^2x L_{\tilde{L}}^{(t)}=\int d^2x \sqrt{-\hat{g}}\mathcal{L}_{\tilde{L}}^{(t)}=\sum_{n \ge 0} t^{n} S_{\tilde{L}}^{(n)}=\int d^2x \sum_{n\ge 0}t^{n}L_{\tilde{L}}^{(n)}=\int d^2x \sqrt{-\hat{g}}\sum_{n\ge 0}t^{n}\mathcal{L}_{\tilde{L}}^{(n)}.
\end{equation} 
We also expand the deformed stress tensor as $t^{\tilde{L}}_{\mu\nu}=\sum_{n\ge0} t^{n} t^{\tilde{L}(n)}_{\mu\nu}$, with $t^{\tilde{L}(n)}_{\mu\nu}=\frac{4\pi}{\sqrt{-\hat{g}}}\frac{\delta S^{(n)}_{\tilde{L}}}{\delta \hat{g}^{\mn}}$. The flow of the action implies the following flow for the Lagrangian density
\begin{equation}
	\frac{d}{dt}\mathcal{L}_{\tilde{L}}^{(t)}=\frac{1}{4\pi}e^{-\sigma} \mathcal O^{\tilde L}_{T\bar T,\hat g}.
\end{equation}
This means that, to first order, we have
\begin{equation}
	\mathcal{L}_{\tilde{L}}^{(1)}=\frac{1}{4\pi}e^{-\sigma} \mathcal O^{\tilde L (0)}_{T\bar T,\hat g},
\end{equation}
with $\mathcal O^{\tilde L (0)}_{T\bar T,\hat g} \equiv t_{\tilde L (0)}^\mn t_{\tilde L}^{\ab (0)} \hat g_{\alpha\mu} \hat g_{\beta\nu} - (t_{\tilde L}^{\mn (0)}\hat g_\mn)^2$. We find
\begin{equation}
	t^{\mu (0)}_{\tilde{L} \mu}=\frac{c}{12 \pi}\hat{\Box}\sigma
\end{equation}
and
\begin{equation}
	\mathcal{O}^{\tilde L (0)}_{T\bar T,\hat g}=\left(\frac{c}{24}\right)^2\left(\frac{1}{2}(\partial \sigma)^4+4 \partial^{\mu}\partial^{\nu}\sigma\partial_{\mu}\partial_{\nu}\sigma-4 (\hat{\Box}\sigma)^2 -4 \partial^{\mu}\sigma\partial^{\nu}\sigma \partial_{\mu}\partial_{\nu}\sigma
	+2\hat{\Box}\sigma\partial^{\mu}\sigma\partial_{\mu}\sigma\right).
\end{equation}
Therefore, the $T\bar{T}$-like deformed Liouville takes the form 
\begin{align}
	S_{\tilde{L}}=S_L+\frac{t}{16\pi}\left(\frac{c}{12}\right)^2 \int d^2x \sqrt{-\hat{g}}\, e^{-\sigma}\Bigg(&\frac{1}{2}(\partial \sigma)^4+4 \partial^{\mu}\partial^{\nu}\sigma\partial_{\mu}\partial_{\nu}\sigma-4 (\hat{\Box}\sigma)^2 \nonumber
	\\
	&-4 \partial^{\mu}\sigma\partial^{\nu}\sigma \partial_{\mu}\partial_{\nu}\sigma
	+2\hat{\Box}\sigma\partial^{\mu}\sigma\partial_{\mu}\sigma\Bigg)+\mathcal{O}(t^2), \label{StildeLsigmat1app}
\end{align}
where we have shown the first non-trivial order explicitly, while the rest can be computed systematically in the expansion. We can check that
\begin{equation}
	t_{\mn}^{\tilde{L}(1)}= \left(\frac{c}{24}\right)^2  e^{-\sigma}\left(-\frac{1}{2}\hat{g}_{\mn} B +2 B_{\mn}\right),
\end{equation}
where we defined $\mathcal{O}^{\tilde L (0)}_{T\bar T,\hat g}=\left(\frac{c}{24}\right)^2 B$, and $B_{\mu\nu}$ obeys $B=\hat{g}^{\mn}B_{\mn}$. Therefore, its trace is
\begin{equation}
	\hat{g}^{\mn}t_{\mn}^{\tilde{L}(1)}= \left(\frac{c}{24}\right)^2  e^{-\sigma}B.
\end{equation}
With this, we can check explicitly that the trace flow equation is satisfied to linear order
\begin{equation}
	\hat{g}^{\mn}t_{\mn}^{\tilde{L}(0)}+t\hat{g}^{\mn}t_{\mn}^{\tilde{L}(1)}=\frac{c}{12}\hat{\Box}\sigma+t e^{-\sigma}\mathcal{O}^{\tilde L (0)}_{T\bar T,\hat g}.
\end{equation}
Integrating by parts, the $T\bar{T}$-like deformed Liouville can be rewritten in the useful forms
\begin{align}
	S_{\tilde{L}}=S_L+\frac{t}{16\pi}\left(\frac{c}{12}\right)^2 \int d^2x \sqrt{-\hat{g}}\, e^{-\sigma}\Bigg(&\frac{1}{2}(\partial \sigma)^4-2\hat{\Box}\sigma(\partial \sigma)^2\Bigg)+\mathcal{O}(t^2)  \label{StildeLsigmat1app2}
\end{align}
and\footnote{It is clear that $\mathcal{O}(t^2)$ or higher contributions will be at least of order $\mathcal{O}(\sigma^3)$.}
\begin{equation}
	S_{\tilde{L}}=S_L+\mathcal{O}(\sigma^3). \label{StildeLsigmat1app1}
\end{equation}

It is instructive to consider simplifying the seed theory by neglecting the curvature coupling, such that the undeformed theory is simply that of the free boson $\sigma$
\ali{
	S_{\tilde{F}}^{(0)} &= \frac{c}{48\pi} \int d^2 x \sqrt{-\hat g}  \frac{1}{2} \hat g^\mn \p_\mu \sigma \p_\nu \sigma,  
}
with stress tensor
\ali{
	t_\mn^{\tilde{F} (0)} = 
	-\frac{c}{24} \left(-\p_\mu \sigma \p_\nu \sigma + \frac{1}{2} \hat g_\mn   \hat g^\ab \p_\al \sigma \p_\beta \sigma   \right). \label{tLappfb} 
}
That is, we are looking for the solution $S_{\tilde{F}}\equiv S^{(t)}_{\tilde{F}}$ of 
\ali{
	\frac{d}{dt} S_{\tilde{F}}^{(t)} &= \frac{1}{4\pi} \int d^2 x \sqrt{-\hat g} \,  e^{-\sigma} \mathcal O^{\tilde{F}}_{T\bar T,\hat g}, \label{actionflowappfb} 
}
corresponding to solutions of the simplified trace flow equation
\ali{
	t_{\tilde{F} \, \mu}^\mu= t \, e^{-\sigma} \mathcal O^{\tilde{F}}_{T\bar T,\hat g}, \qquad \mathcal O^{\tilde{F}}_{T\bar T,\hat g} \equiv t_{\tilde{F}}^\mn t_{\tilde{F}}^\ab \hat g_{\alpha\mu} \hat g_{\beta\nu} - (t_{\tilde{F}}^\mn \hat g_\mn)^2, \label{traceflowliouappfb} 
}
with deformed stress tensor $t_{\mn}^{\tilde{F}}=\frac{4\pi}{\sqrt{-\hat{g}}}\frac{\delta S_{\tilde{F}}}{\delta \hat{g}^{\mn}}$. Employing the analogous notation for the expansions that we have used before, we have
\begin{equation}
	\frac{d}{dt}\mathcal{L}^{(t)}_{\tilde{F}}=\frac{1}{4\pi}e^{-\sigma} \mathcal O^{\tilde F}_{T\bar T,\hat g} 
\end{equation}
and thus
\begin{equation}
	\mathcal{L}^{(1)}_{\tilde{F}}=\frac{1}{4\pi}e^{-\sigma} \mathcal O^{\tilde F (0)}_{T\bar T,\hat g},
\end{equation}
with $\mathcal O^{\tilde{F} (0)}_{T\bar T,\hat g} \equiv t_{\tilde{F} (0)}^\mn t_{\tilde{F}}^{\ab (0)} \hat g_{\alpha\mu} \hat g_{\beta\nu} - (t_{\tilde{F}}^{\mn (0)}\hat g_\mn)^2$. We find
\begin{equation}
	t^{\mu (0)}_{\tilde{F} \mu}=0
\end{equation}
and
\begin{equation}
	\mathcal{O}^{\tilde{F} (0)}_{T\bar T,\hat g}=\frac{1}{2}\left(\frac{c}{24}\right)^2(\partial \sigma)^4.
\end{equation}
Therefore, the $T\bar{T}$-like deformed free boson takes the form 
\begin{align}
	S^{(t)}_{\tilde{F}}=S^{(0)}_{\tilde{F}}+\frac{t}{16\pi}\left(\frac{c}{12}\right)^2 \int d^2x \sqrt{-\hat{g}}\, e^{-\sigma}\frac{1}{2}(\partial \sigma)^4+\mathcal{O}(t^2), \label{StildeLsigmat1appfb}
\end{align}
where we have shown the first non-trivial order explicitly, while the rest can be computed systematically in the expansion. In this case, it is easy to obtain a closed form, either by
finding a recursive relation and resumming it to all orders \cite{Cavaglia:2016oda}, or by 
obtaining the associated Burger's equation with well-known solution \cite{Bonelli:2018kik}. We find
\ali{
	S^{(t)}_{\tilde{F}} &= \frac{1}{4\pi} \int d^2 x \sqrt{-\hat g} \,\, e^{\sigma} \, \frac{1 - \sqrt{1 -  \, \frac{t \, c}{12}  e^{-\sigma} \hat g^\mn \p_\mu \sigma \p_\nu \sigma}}{t}  \label{StildeLsigmaapp}
} 
with corresponding deformed stress tensor
\begin{equation}
	t^{\tilde{F}}_{\mn}=\frac{A_{\mn}}{2\sqrt{1-t e^{-\sigma}A}}+g_{\mn}e^{\sigma}\frac{\sqrt{1-t e^{-\sigma}A}-1}{2t}
\end{equation}
here given in terms of $A_{\mu\nu}=\frac{c}{12}\partial_{\mu}\sigma \partial_{\nu}\sigma$ and $A=\hat{g}^{\mn}A_{\mn}$. The trace of the deformed stress tensor is
\begin{equation}
	\hat{g}^{\mn}t^{\tilde{F}}_{\mn}=\frac{2 e^{\sigma}(1-\sqrt{1-te^{-\sigma}A})-At}{2t\sqrt{1-te^{-\sigma}A}}
\end{equation}
while the $T\bar{T}$ operator evaluates to
\begin{equation}
	\mathcal{O}_{T\bar{T},\hat{g}}^{\tilde{F}}=\frac{2e^{2\sigma} (1-\sqrt{1-te^{-\sigma}A})-e^{\sigma}At}{2t^2\sqrt{1-te^{-\sigma}A}},
\end{equation}
thus confirming that $S_{\tilde{F}}$ is indeed an exact solution to both $\frac{d}{dt} S_{\tilde{F}}^{(t)} = \frac{1}{4\pi} \int d^2 x \sqrt{-\hat g} \,  e^{-\sigma} \mathcal O^{\tilde{F}}_{T\bar T,\hat g}$ and  $	t_{\tilde{F} \, \mu}^\mu= t \, e^{-\sigma} \mathcal O^{\tilde{F}}_{T\bar T,\hat g}$. 

\bibliographystyle{JHEP}
\bibliography{referencesTTbarBWDraft}
	
\end{document}